\documentclass[%
 reprint,
 amsmath,amssymb,
 aps,
pra,
]{revtex4-1}

\usepackage{graphicx}
\usepackage{dcolumn}
\usepackage{bm}
\usepackage{cancel,amsmath,amsfonts,subfigure,mathtools}
\usepackage[mathscr]{eucal}

\graphicspath{{figs/}{figs/fig1/}{figs/fig2/}{figs/fig3/}{figs/fig4/}{figs/fig5/}{figs/fig6/}{figs/fig7/}{figs/fig8/}{figs/fig9/}{figs/fig10/}{figs/fig11/}{figs/fig12/}}

\renewcommand{\vec}[1]{\mathbf{\boldsymbol{#1}}}

\begin{document}


\title{Optical selection rules of zigzag graphene nanoribbons}

\author{V. A. Saroka}
\email{v.saroka@exeter.ac.uk}
\affiliation{School of Physics, University of Exeter, Stocker Road, Exeter EX4 4QL, United Kingdom}
\affiliation{Institute for Nuclear Problems, Belarusian State University, Bobruiskaya 11, 220030 Minsk, Belarus}
\author{M. V. Shuba}
\affiliation{Institute for Nuclear Problems, Belarusian State University, Bobruiskaya 11, 220030 Minsk, Belarus}
\author{M. E. Portnoi}
\affiliation{School of Physics, University of Exeter, Stocker Road, Exeter EX4 4QL, United Kingdom}



\date{\today}

\begin{abstract}
We present an analytical tight-binding theory of the optical properties of graphene nanoribbons with zigzag edges. Applying the transfer matrix technique to the nearest-neighbor tight-binding Hamiltonian, we derive analytical expressions for electron wave functions and optical transition matrix elements for incident light polarized along the structure axis. It follows from the obtained results that optical selection rules result from the wave function parity factor $(-1)^J$, where $J$ is the band number. These selection rules are that $\Delta J$ is odd for transitions between valence and conduction subbands and that  $\Delta J$ is even for transitions between only valence (conduction) subbands. 
Although these selection rules are different from those in armchair carbon nanotubes, there is a hidden correlation between absorption spectra of the two structures that should allow one to use them interchangeably in some applications.
The correlation originates from the fact that van Hove singularities in the tubes are centered between those in the ribbons if the ribbon's width is about a half of the tube's circumference. The analysis of the matrix elements’ dependence on the electron wave vector for narrow ribbons shows a smooth non-singular behavior at the Dirac points and the points where the bulk states meet the edge states.
\end{abstract}

\maketitle


\section{\label{sec:Introduction}Introduction}
Graphene nanoribbons with zigzag edges are quasi-one-dimensional nanostructures based on graphene~\cite{Novoselov2004} that are famous for their edges states. These states were theoretically predicted for ribbons with the zigzag edge geometry by Fujita~\cite{Fujita1996} and for a slightly modified zigzag geometry by Klein~\cite{Klein1994}, although the history could be dated back to the pioneering works on polymers~\cite{Kivelson1983,Tanaka1987}. Since then, edge states in zigzag ribbons have been attracting much attention from the scientific community~\cite{Nakada1996,Miyamoto1999,Wakabayashi1999,Lin2000,Wakabayashi2001,Brey2006,Sasaki2006b,Chang2006,Son2006,Son2006a,Kohmoto2007,Yang2007a,Malysheva2008,Onipko2008,Wakabayashi2010,Dutta2010,Chung2011,Wakabayashi2012,Deng2013,Deng2014,Chung2016}, because such peculiar localization of the states at the edge of the ribbon should result in the edge magnetization due to the electron-electron interaction. Although the effect was proved to be sound against an edge disorder~\cite{Nakada1996}, such an edge magnetization had not been experimentally confirmed until quite recently~\cite{Magda2014a}. A fresh surge of interest to physics of zigzag nanoribbons is expected due to the recent synthesis of zigzag ribbons with atomically smooth edges~\cite{Ruffieux2016} and a rapid development of the self-assembling technique~\cite{Schulz2017}.

The edge states in zigzag ribbons have been predicted to be important in transport~\cite{Deng2013,Deng2014,Luck2015}, electromagnetic~\cite{Brey2007}, and optical properties~\cite{Lin2000,Chang2006,Hsu2007}. Although considerable attention has been given to zigzag ribbons' optical properties~\cite{Lin2000,Chang2006,Hsu2007,Sasaki2009,Duan2010,Sasaki2011,Gundra2011,Chung2011,Sanders2012,Sanders2013,Zhu2013a,Chung2016}, including many-body effects~\cite{Yang2007,Prezzi2008,Gundra2011}, the effect of external fields~\cite{Chang2006,Gundra2011}, curvature~\cite{Chung2016}, wave function overlapping integrals~\cite{Sanders2012,Sanders2013}, the finite length effect~\cite{Berahman2011}, and the role of unit cell symmetry~\cite{Gundra2011a}, a number of problems have not been covered yet. In particular, it is known that the optical matrix element of graphene is anisotropic at the Dirac point~\cite{Gruneis2003,Saito2004} due to the topological singularity inherited from the wave functions~\cite{Ando2002,Ando2005}. However, the fate of this singularity in the presence of the edge states, i.e., in zigzag nanoribbons, has not been investigated. This requires analysis of the optical transition matrix element dependence on the electron wave vector, in contrast to the usual analysis limited solely to the selection rules.

It was obtained numerically by Hsu and Reichl that the optical selection rules for zigzag ribbons are different from those in armchair carbon nanotubes~\cite{Hsu2007}. By matching the number of atoms in the unit cell of a zigzag ribbon and an armchair tube, it was demonstrated that the optical absorption spectra of both structures are qualitatively different~\cite{Hsu2007}. However, a comparison of these structures based on the matching of their boundary conditions, similar to what has been accomplished for the band structures~\cite{White2007a} and optical matrix elements~\cite{Portnoi2015} of armchair graphene nanoribbons and zigzag carbon nanotubes, has not been reported yet.

The distinctive selection rules of zigzag graphene nanoribbons were noticed as early as 2000 by Lin and Shyu~\cite{Lin2000}. This remarkable and counter-intuitive result, especially when compared to the optical selection rules of carbon nanotubes~\cite{Ajiki1994,Gruneis2003,Jiang2004,Goupalov2005,Malic2006,Zarifi2009}, was obtained numerically and followed by a few attempts to provide an analytical explanation~\cite{Chung2011,Sasaki2011}. 

Within the nearest-neighbor approximation of the $\pi$-orbital tight-binding model the optical selection rules for graphene nanoribbons with zigzag edges is a result of the wave function parity factor $(-1)^J$, where $J$ numbers conduction (valence) subbands. This factor has been obtained numerically as a connector of wave function components without explicit expressions for the wave functions being presented~\cite{Chung2011}. Concurrently, the factor $(-1)^J$, responsible for the optical selection rules, is missing in some papers providing explicit expressions for the electron wave functions (see Appendix of Ref.~\onlinecite{Wakabayashi2010}). Although it emerged occasionally in later works dealing with the transport and magnetic properties of the ribbons~\cite{Wakabayashi2012,Deng2014}, its important role was not emphasized and its origin remains somewhat obscure. At the same time, Sasaki and co-workers obtained the optical matrix elements which, although providing the same selections rules, are very different from those in Ref.~\onlinecite{Chung2011}. Moreover, despite being reduced to the low-energy limit around the Dirac point, the matrix elements in Ref.~\onlinecite{Sasaki2011} remain strikingly cumbersome.

It is the purpose of the present paper to demonstrate a simple way of obtaining analytical expressions for optical transition matrix elements in the orthogonal tight-binding model. The essence of this work is an analytical refinement of the paper by Chung \emph{et al.}~\cite{Chung2011}, which provides an alternative explanation of the selection rules to that given in terms of pseudospin~\cite{Sasaki2011}. However, we do not simply derive analytically the results of the study~\cite{Chung2011} showing their relation to the zigzag ribbon boundary condition and secular equation, but extend the approach to the transitions between conduction (valence) subbands considered by Sasaki \emph{et al.}~\cite{Sasaki2011}. Unlike both mentioned studies, we go beyond a ``single point" consideration of the optical matrix elements and analyze the matrix elements as functions of the electron wave vector.
The presence of possible singularities in these dependencies at $k=2 \pi /3$, corresponding to the Dirac point, and at the transition point $k_t$, where the edge states meet bulk states, is in the scope of our study. It is also the purpose of this paper to investigate relations between zigzag ribbons' and armchair nanotubes' optical properties by matching their boundary conditions in lieu of matching the number of atoms in the unit cells as was done by Hsu and Reichl~\cite{Hsu2007}.

This paper is organized as follows. In Sec.~\ref{sec:AnalyticalTightBindingModel} we present the tight-binding Hamiltonian and solve its eigenproblem by the transfer matrix method, following the original paper by Klein~\cite{Klein1994}, in this section, many analogies can be drawn with the treatment of finite length zigzag carbon nanotubes~\cite{Compernolle2003}; optical transition matrix elements are derived within the so-called gradient (effective mass) approximation, and optical selection rules are obtained. The analytical results are discussed and supplemented by a numerical study in Sec.~\ref{sec:Discussion}. Finally, the summary is provided in Section~\ref{sec:Conclusions}. We relegate to the appendixes some technical details on ribbon wave functions and supplementary results on matching periodic and ``hard wall" boundary conditions.

\section{\label{sec:AnalyticalTightBindingModel} Analytical tight-binding model}

\subsection{Hamiltonian eigenproblem}
\begin{figure}[tbhp]
    \includegraphics[scale=0.45]{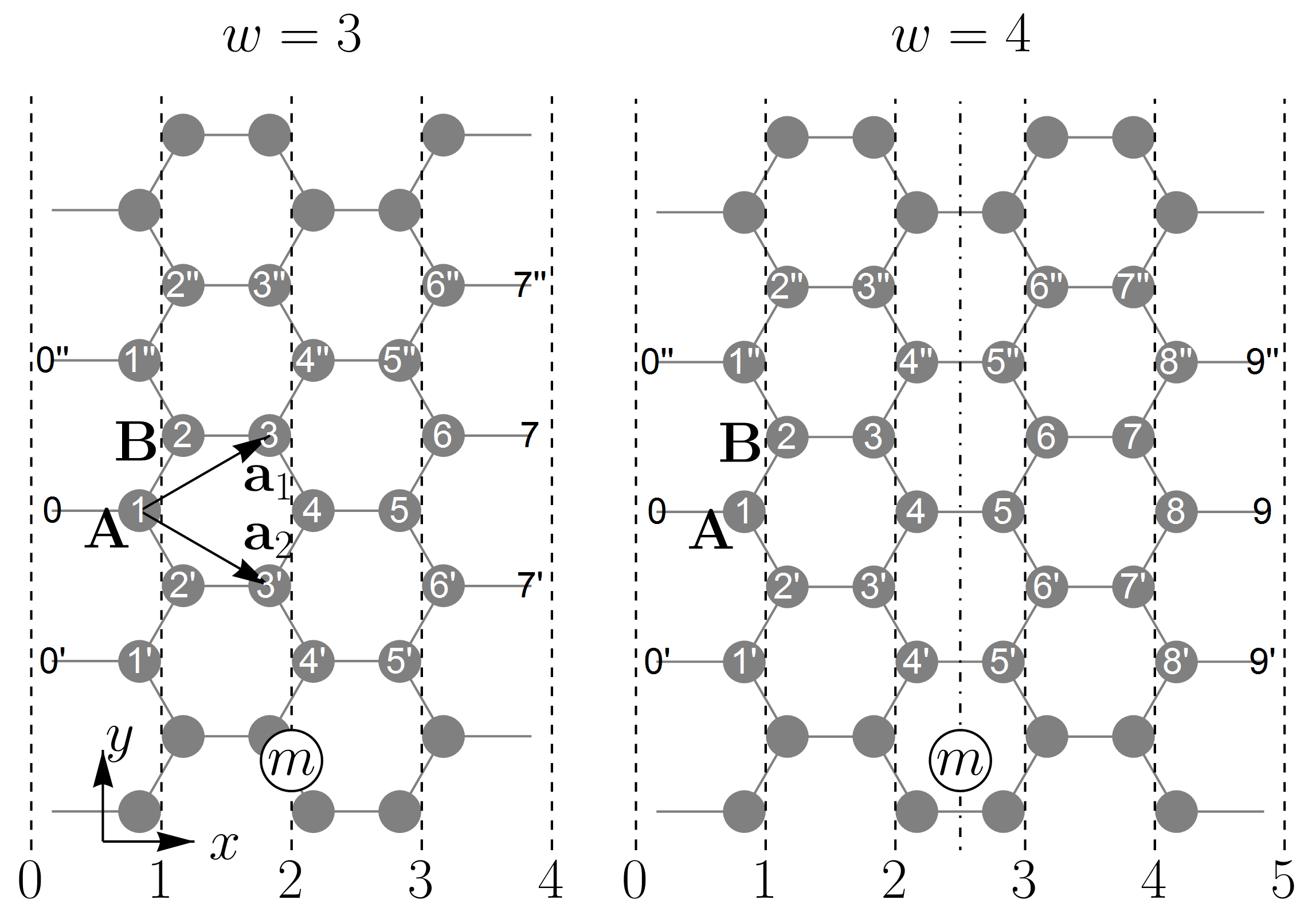}
	\caption{\label{fig:ZGNRAtomicStructureAndNumberingMFLin} The atomic structure of zigzag ribbons consisting of $w=3$ and $4$ zigzag chains. The carbon atoms are numbered within the ribbon unit cells. The two outermost sites, where the electron wave function vanishes, are labeled by black numbers. The graphene lattice primitive translations $\vec{a}_1$ and $\vec{a}_2$ are shown along with the two nonequivalent atoms from the  A and B sublattices forming the honeycomb lattice of graphene. The positions of zigzag chains, including auxiliary ones, where the electron wave function vanishes, are marked by dashed lines. $m$ labels the dashed dotted line of the mirror symmetry for even $w$ and the ribbon center for odd $w$.}
\end{figure}

Let us consider a zigzag ribbon within the tight-binding model, which is the orthogonal $\pi$-orbital model taking into account only nearest-neighbor hopping integrals. The atomic structure of a graphene nanoribbon with zigzag edges is presented in Fig.~ \ref{fig:ZGNRAtomicStructureAndNumberingMFLin}.  A ribbon with a particular width can be addressed by index $w$, numbering trans-polyacetylene chains --- so-called ``zigzag" chains.
For such a ribbon, the tight-binding Hamiltonian can be constructed in the usual way by putting $k_x \rightarrow 0$, where $k_x$ is the transverse component of the electron wave vector. We avoid the procedure described by Klein~\cite{Klein1994}, since it results in a Hamiltonian for which concerns were raised by Gundra and Shukla~\cite{Gundra2011a}. Thus, for the ribbon with $w=2$, it reads
\begin{equation}
H = \label{eq:ZGNRProblemHamiltonian}
\left(\begin{array}{cccccc}
0 & \gamma q & 0 & 0  \\
\gamma q & 0 & \gamma & 0  \\
0 & \gamma & 0 & \gamma q  \\
0 & 0 & \gamma q & 0 
\end{array}\right)
\end{equation}
where $\gamma$ is the hopping integral and $q = 2 \cos (k/2)$ with $k=k_y a$ being the dimensionless electron wave vector and $a=|\vec{a}_1| = |\vec{a}_2| = 2.46$~{\AA} being the graphene lattice constant. 
The Hamiltonian $H$ has a tridiagonal structure, therefore its eigenproblem can be solved by the transfer matrix method, which is a general mathematical approach for analytical treatment of tridiagonal and triblock diagonal matrix eigenproblems~\cite{Molinari1997}. This approach was developed and widely used for investigation of one-dimensional systems~\cite{Kerner1954,Schmidt1957,Hori1957a,Matsuda1962a}. An alternative approach may be based on continuants, which also have been using for the investigation of conjugated $\pi$-carbons such as polyenes and aromatic molecules~\cite{LennardJones1937,Coulson1937} and carbon nanotubes~\cite{Mestechkin2005} (see also Refs.~\onlinecite{Rutherford1948,Rutherford1952,BookMuir1960}).

We use $H$  to derive the relations between the eigenvector components presented in the paper by Chung \emph{et al.}~\cite{Chung2011}. In particular, we pay special attention to the origin of the $(-1)^J$  factor and its relation to the eigenstate parity. In the rest of this section, we solve the eigenproblem for $H$.

\subsubsection{Eigenvalues: proper energy}
In this part of the section, we find eigenvalues by the transfer matrix method~\cite{Kerner1954,Schmidt1957,Hori1957a,Matsuda1962a}. The eigenproblem for the Hamiltonian given by Eq.~\eqref{eq:ZGNRProblemHamiltonian} can be written as follows:
\begin{align}
\label{eq:ZGNRProblemSystemOfEquations}
c_{j-1} \gamma - c_j E + c_{j+1}  \gamma q &= 0, &  j&= 2 p -1 \, ; \\ \nonumber
c_{j-1} \gamma q - c_j E + c_{j+1} \gamma &= 0, &  j&= 2 p \, ;
\end{align}
where $p=1, \dots, w$, $w=N/2$, and $N$ is the number of atoms in the ribbon unit cell. Each of the equations above can be rewritten in the transfer matrix form~\cite{Schmidt1957}:
\begin{align}
\label{eq:ZGNRProblemTransferMatrixFormOfSystemOfEquations}
\left(\begin{array}{c}
c_j \vphantom{-\dfrac{1}{q}} \\
c_{j+1} \vphantom{-\dfrac{1}{q}}
\end{array}
\right) &=\left(
\begin{array}{cc}
\phantom{-}0 \vphantom{-\dfrac{1}{q}} & \phantom{-}1 \vphantom{-\dfrac{1}{q}} \\
-\dfrac{1}{q} & \phantom{-}\dfrac{\alpha}{q}
\end{array}\right)
\left(
\begin{array}{c}
c_{j-1} \vphantom{-\dfrac{1}{q}} \\
c_{j} \vphantom{-\dfrac{1}{q}}
\end{array}
\right), & j&=2p-1\, ; \\ \nonumber
\left(\begin{array}{c}
c_j \\
c_{j+1}
\end{array}
\right) &= \left(
\begin{array}{cc}
\phantom{-}0 & \phantom{-}1 \\
-q & \phantom{-}\alpha
\end{array}
\right)
\left(
\begin{array}{c}
c_{j-1} \\
c_{j}
\end{array}
\right), & j&=2p \, ;
\end{align}
where $\alpha = E/\gamma$.
Introducing
\begin{align}
T_1 &= \left(
\begin{array}{cc}
\phantom{-}0 & \phantom{-}1 \\
-\dfrac{1}{q} & \phantom{-}\dfrac{\alpha}{q}
\end{array}
\right)\, , &
T_2 &= \left(
\begin{array}{cc}
\phantom{-}0 & \phantom{-}1 \\
-q & \phantom{-} \alpha
\end{array}
\right) \, ,
\end{align}
and substituting $j$ into (\ref{eq:ZGNRProblemTransferMatrixFormOfSystemOfEquations}) yield
\begin{align}
\label{eq:ZGNRProblemTransferMatrixFormOfSystemOfEquations1}
\left(\begin{array}{c}
c_{2p-1} \\
c_{2p}
\end{array}
\right) &= T_1
\left(
\begin{array}{c}
c_{2p-2} \\
c_{2p-1}
\end{array}
\right) \, , \\ \nonumber
\left(\begin{array}{c}
c_{2p} \\
c_{2p+1}
\end{array}
\right) &= T_2
\left(
\begin{array}{c}
c_{2p-1} \\
c_{2p}
\end{array}
\right) \, ,
\end{align}
whence the following recursive relation can be readily noticed:
\begin{align}
\label{eq:ZGNRProblemTransferMatrixFormOfSystemOfEquations2}
\left(\begin{array}{c}
c_{2p} \\
c_{2p+1}
\end{array}
\right) &= T_2
T_1
\left(
\begin{array}{c}
c_{2p-2} \\
c_{2p-1}
\end{array}
\right) \, ,
\end{align}
and the following transfer matrix equation can be obtained:
\begin{align}
\label{eq:ZGNRProblemTransferMatrixEquation}
C_{2p+1} &= \left( \begin{array}{c}
c_{2p} \\
c_{2p+1}
\end{array} \right) = T^{p} C_1 \, .
\end{align}
Thus, the transfer matrix in question is
\begin{equation}
T = T_2\, T_1 = \left(\begin{array}{cc}
- \dfrac{1}{q} & \dfrac{\alpha}{q} \\
-\dfrac{\alpha}{q} & \dfrac{\alpha^2- q^{2}}{q}
\end{array}\right) \, .
\end{equation}
The characteristic equation for finding the eigenvalues of $T$, $\mbox{det}\left(T-\lambda I\right) =0$, is a quadratic one: 
\begin{equation}
\lambda^2 +\left(\dfrac{1}{q} + q - \dfrac{\alpha^2}{q}\right) \lambda + 1 = 0 \, .
\end{equation}
This equation has the following solution:
\begin{equation}
\lambda_{1,2} =   A \pm \sqrt{A^2 - 1} \, ,
\end{equation}
where
\begin{equation}
\label{eq:ZGNRProblemCosTheta}
A = \dfrac{\alpha^2 - q^{2} -1}{2 q} = - \cos \theta \, .
\end{equation}
A new variable $\theta$ has been introduced above to reduce the eigenvalues $\lambda_{1,2}$ to the complex exponent form, which is favourable for further calculations:
\begin{equation}
\label{eq:ZGNRProblemTransferMatrixEigenValues}
\lambda_{1,2} = - e^{\mp i \theta } \, ,
\end{equation}
where the upper (lower) sign is used for $\lambda_1$ ($\lambda_2$). We must note that another choice of variable $\theta$, i.e., $A=\cos \theta$, is also possible, but it results in the inverse numbering of the proper energy branches. The minus sign is a better choice because it allows one to avoid a change of the lowest (highest) conduction (valence) subband index when the ribbon width increases.

Equation (\ref{eq:ZGNRProblemCosTheta}) allows one to express the proper energy in terms of $\theta$ and $q$:
\begin{equation}
\label{eq:ZGNRProblemProperEnergy0}
\alpha = \dfrac{E}{\gamma} = \pm \sqrt{ q^{2} - 2 q \cos \theta + 1} \, .
\end{equation}
Taking into account that $q = 2 \cos (k/2)$, for the proper energy, we obtain
\begin{equation}
\label{eq:ZGNRProblemProperEnergy}
E = \pm \gamma \sqrt{ 4 \cos^2 \dfrac{k}{2} - 4 \cos \dfrac{k}{2} \cos \theta + 1} \, ,
\end{equation}
where $\theta$ is to be found from the secular equation for the fixed ends boundary condition as in the case of a finite atomic chain~\cite{Schmidt1957,Hori1957a,Matsuda1962a}. The physical interpretation of the parameter $\theta$ is to be given further. We note that Eq.~\eqref{eq:ZGNRProblemProperEnergy} has similar form not only to the graphene energy band structure~\cite{Wallace1947,Saito1992,SaitoBook1998} but also to the eigenenergies of the finite length zigzag carbon nanotubes~\cite{Compernolle2003} [cf. with Eq.~(32) therein].

\subsubsection{Secular equation}
For the fixed end boundary condition, which, in the context of the electronic properties being considered, is better referred to as the ``hard wall" boundary condition, the general form of the secular equation is $(T^w)_{22}=0${}~\cite{Matsuda1962a}. This equation can be obtained by imposing the constraint $c_{0}=c_{N+1}=0$ on Eq.~\eqref{eq:ZGNRProblemTransferMatrixEquation}, where $p=w$, which physically means the vanishing of the tight-binding electron wave functions on sites $0$ and $N+1$, or equivalently on zigzag chains $0$ and $w+1$ as illustrated in Fig.~\ref{fig:ZGNRAtomicStructureAndNumberingMFLin}. Hence, for the secular equation, the $w$-th power of the transfer matrix $T$ is needed. The simplest way of calculating $T^w$ is $T^w=S \Lambda^w S^{-1}$, where $\Lambda$ is the diagonal form of $T$ and $S$ is the matrix making the transformation to a new basis in which $T$ is diagonal. The eigenvalues of $T$ are given by Eq.~\eqref{eq:ZGNRProblemTransferMatrixEigenValues}, therefore, $\Lambda$ can be easily written down. Concurrently, the $S$ matrix can be constructed from eigenvectors of $T$ written in columns. By setting the first components of the vectors to be equal to unity, one can reduce them to 
\begin{align}
\label{eq:ZGNRProblemV1andV2}
V_1 &= \left(\begin{array}{c}
1 \\
\xi_1
\end{array}\right), &
V_2 &= \left(\begin{array}{c}
1 \\
\xi_2
\end{array}\right)\, ,
\end{align}
where the following notation is used:
\begin{equation}
\label{eq:ZGNRProblemXi12}
\xi_{1,2} = \dfrac{1 + q \lambda_{1,2}}{\alpha}\, .
\end{equation}
Then the matrix $S$ and its inverse matrix $S^{-1}$ can be written as follows:
\begin{align}
\label{eq:ZGNRProblemSinv}
S &= \left(\begin{array}{cc}
1 & 1 \\
\xi_1 & \xi_2
\end{array}\right), & 
S^{-1} &= \dfrac{1}{\xi_2 - \xi_1}\left(\begin{array}{cc}
\xi_2 & -1 \\
-\xi_1 & \phantom{-}1
\end{array}\right) \, .
\end{align}
Expressions~\eqref{eq:ZGNRProblemSinv} are of the same form as in the atomic ring problem~\cite{Hori1957a}. Using (\ref{eq:ZGNRProblemSinv}), the $T^w$ calculation yields 
\begin{equation}
\label{eq:ZGNRProblemTw}
T^w = \dfrac{1}{\xi_2 - \xi_1} \left(\begin{array}{cc}
\xi_2 \lambda_1^w - \xi_1 \lambda_2^w & \lambda_2^w - \lambda_1^w \\
\xi_1 \xi_2 \left(\lambda_1^w - \lambda_2^w \right) & \xi_2 \lambda_2^w - \xi_1 \lambda_1^w
\end{array}\right) \, .
\end{equation}
Now by the aid of (\ref{eq:ZGNRProblemXi12}) and (\ref{eq:ZGNRProblemTransferMatrixEigenValues}) from (\ref{eq:ZGNRProblemTw}), we can find the explicit form of the secular equation for $\theta$: 
\begin{equation}
\label{eq:ZGNRProblemQuantizationCondition}
 \sin w  \theta -  2 \cos \dfrac{k}{2} \, \sin \left[(w+1)\theta \right]  = 0 \, .
\end{equation}
The equation above is very much like that analyzed by Klein~\cite{Klein1994} for so-called ``bearded" zigzag ribbons, therefore, the same basic analysis can be carried out.

\begin{figure}
    \includegraphics[scale=0.45]{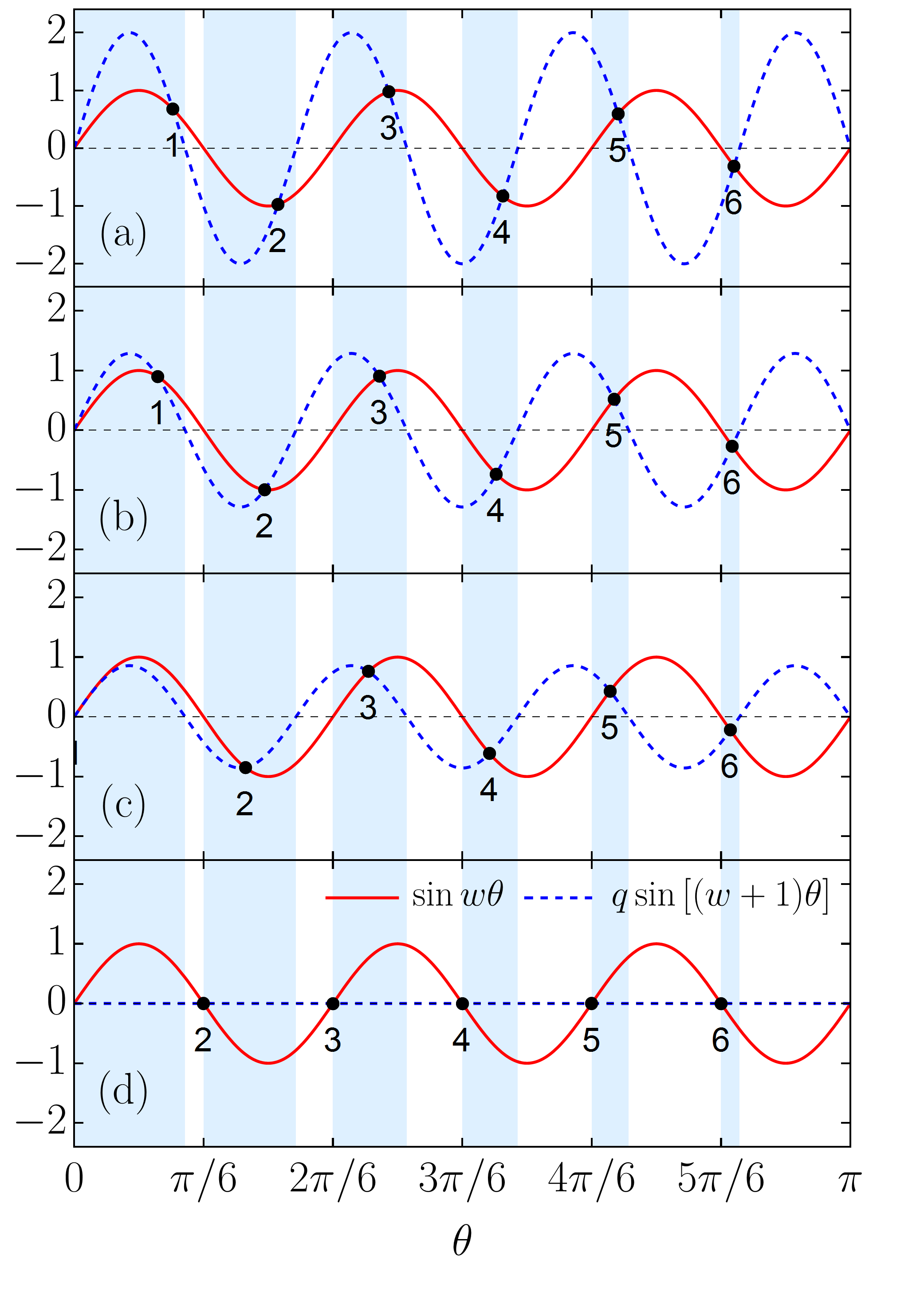}
	\caption{Solutions of the secular equation~\eqref{eq:ZGNRProblemQuantizationCondition} for zigzag graphene nanoribbon with $w=6$ and the following values of the parameter $q = 2 \cos (k/2)$: (a) $2$, (b) $3w/2(w+1)$, (c) $w/(w+1)$, and (d) $0$. The light blue shading signifies the $\theta$ intervals to which the secular equation solutions are confined for $q$'s ranging from $0$ to $\infty$.}
	\label{fig:ZGNRProblemTBmodelQC}
\end{figure}

As can be seen from Fig.~\ref{fig:ZGNRProblemTBmodelQC}, all nonequivalent solutions of Eq.~\eqref{eq:ZGNRProblemQuantizationCondition} reside in the interval $\theta \in (0,\pi)$. When the slope of $q \sin[(w+1) \theta]$ at $\theta=0$  is greater than that of $\sin w\theta$ , i.e.,
$(q \sin[(w+1) \theta])^{\prime}_{\theta=0} > (\sin w \theta)^{\prime}_{\theta=0} \Rightarrow 2 \cos (k/2) > w/(w+1)$, there are $w$ different solutions in the interval, which give $2w$ branches of the proper energy~\eqref{eq:ZGNRProblemProperEnergy}. This is indicated in Figs.~\ref{fig:ZGNRProblemTBmodelQC} (a) and~\ref{fig:ZGNRProblemTBmodelQC} (b). However, as seen from Figs.~\ref{fig:ZGNRProblemTBmodelQC} (c) and~\ref{fig:ZGNRProblemTBmodelQC} (d), when $2 \cos (k/2) \leq w/(w+1)$, one solution is missing and Eq.~\eqref{eq:ZGNRProblemProperEnergy} defines only $2w-2$ branches.
The missing solution can be restored by analytical continuation $\theta = i \beta$, where $\beta$ is a parameter to be found. In this case, the secular equation~\eqref{eq:ZGNRProblemQuantizationCondition} and the proper energy~\eqref{eq:ZGNRProblemProperEnergy} must be modified accordingly by changing trigonometric functions to hyperbolic ones. 

The above introduced parameter $\theta$ ($\beta$) can be interpreted as a transverse component of the electron wave vector and the secular equation~\eqref{eq:ZGNRProblemQuantizationCondition} can be referred to as its quantization condition.

\subsubsection{Eigenvectors: wave functions}
Let us now find eigenvectors of the Hamiltonian given by Eq.~\eqref{eq:ZGNRProblemHamiltonian}. To obtain the eigenvector components, we choose the initial vector $C_1 = (c_0,c_1)$ as a linear combination of the transfer matrix eigenvectors that satisfies the ``hard wall" boundary condition $c_0 = 0$ : $C_1 =  (V_1 - V_2)/(2i)$. It is to be mentioned here that the opposite end boundary condition, $c_{N+1}=0$, is ensured by Eq.~\eqref{eq:ZGNRProblemQuantizationCondition}. The chosen $C_1$ yields 
\begin{align}
C_{2p+1} = T^{p} C_1 &= \dfrac{1}{2 i} \left( \lambda_1^{p} V_1 - \lambda_2^{p} V_2\right) \nonumber\\ 
&= \dfrac{1}{2 i} \left( \begin{array}{c}
\lambda_1^{p} -\lambda_2^{p}  \\
\lambda_1^{p} \xi_1 - \lambda_2^{p} \xi_2
\end{array} \right)
\end{align}
or, equivalently,
\begin{align}
\label{eq:ZGNRProblemEigenVectorComponents}
c_{2p} &= \dfrac{1}{2 i} \left( \lambda_1^{p} - \lambda_2^{p} \right), & p&=1,\dots, w \,  ; \\ \nonumber
c_{2p+1} &= \dfrac{1}{2 i} \left( \lambda_1^{p} \xi_1 - \lambda_2^{p} \xi_2 \right), & &
\end{align}
Substituting~\eqref{eq:ZGNRProblemTransferMatrixEigenValues} and~\eqref{eq:ZGNRProblemXi12} into~(\ref{eq:ZGNRProblemEigenVectorComponents}) and keeping in mind the definition of $\alpha$, one readily obtains
\begin{align}
\label{eq:ZGNRProblemEigenVectorComponent1}
c_{2p} &= (-1)^{p+1} \sin p \theta , \qquad \qquad p=1,\dots, w\, ;  \\
\label{eq:ZGNRProblemEigenVectorComponent2}
c_{2p+1} &= \dfrac{(-1)^{p+1} \gamma}{E} \left\{ \sin p \theta - 2 \cos \dfrac{k}{2} \, \sin \left[ (p+1) \theta \right] \right\} \, . 
\end{align}
It is worth pointing out that for the starting $p=1$ from the equations above one gets components $c_2$ and $c_3$. Although it may seem strange because of the missing $c_1$, this is how it should be for $c_1$ has already been specified by the proper choice of the initial vector $C_1$. 

Equation~\eqref{eq:ZGNRProblemEigenVectorComponent2} can be further simplified (see Appendix~\ref{app:sec:WaveFunctionParityFactor}) so that for the eigenvector components, one has
\begin{align}
\label{eq:ZGNRProblemEigenVectorComponent12}
c^{(j)}_{2p} &= (-1)^{p+1} \sin p \theta_j, \qquad \qquad p=1,\dots, w \, ;   \\
\label{eq:ZGNRProblemEigenVectorComponent22}
c^{(j)}_{2p+1} &= \pm (-1)^{p+1}  (-1)^{j-1}  \sin \left[ (p-w)  \theta_j \right] \, , & &
\end{align}
where we have introduced the index $j$ to number various values of $\theta$, which are solutions of Eq.~\eqref{eq:ZGNRProblemQuantizationCondition}. As one may have noticed the above expressions still have one drawback: $p=1$ defines components $c_2$ and $c_3$, while it would be much more convenient if $p=1$ would instead specify $c_1$ and $c_2$. To obtain desired dependence of the eigenvector components on the variable index, one needs to redefine  in Eq.~\eqref{eq:ZGNRProblemEigenVectorComponent22} the index $p\rightarrow n-1$:
\begin{align*}
c^{(j)}_{2n-1} &= \pm (-1)^{n}  (-1)^j  \sin \left[  (w+1-n) \theta_j \right] \, , & n&=1,\dots, w \, ;  \\
c^{(j)}_{2p} &= (-1)^{p+1} \sin p \theta_j \,,  & p&=1,\dots, w \,;
\end{align*}
and then put $n \rightarrow p$. The latter is permissible since $n$ is a dummy index that can be denoted by any letter. Note that due to the change of the terms order in the sine function one $(-1)$ factor in the coefficient $c^{(j)}_{2p+1}$ above cancels, therefore, $j-1$ in the exponent has been replaced by $j$. Thus, for the Hamiltonian~\eqref{eq:ZGNRProblemHamiltonian}, we end up with the following eigenvectors:
\begin{align}
\label{eq:ZGNRProblemEigenVectorComponents3}
c^{(j)}_{2p-1} &= \mp (-1)^p (-1)^j \sin \left[ (w+1-p) \theta_j \right] \,, & &  \\
\nonumber
c^{(j)}_{2p} &=  (-1)^p \sin p \theta_j  \, , \qquad p= 1,\dots, w \, ,
\end{align}
where we have got rid of $(-1)$ in $c^{(j)}_{2p}$. Since the whole eigenvector $\left| c^{(j)} \right\rangle =  \left( c^{(j)}_1, c^{(j)}_2, \ldots, c^{(j)}_{N}\right)$ can be multiplied by any number, one can choose this number to be $(-1)$. Having multiplied $\left| c^{(j)} \right\rangle$ by $(-1)$, one has to change $\pm$ to $\mp$ in the coefficient $c^{(j)}_{2p+1}$, therefore in Eq.~\eqref{eq:ZGNRProblemEigenVectorComponents3}, the upper ``$-$'' stands now for the conduction band, while the lower ``$+$'' for the valence band. The $(-1)^p$ factor, however, cannot be eliminated in a similar way because it determines the signs of various components differently. Nevertheless, this factor is of no significance, too, for it can be eliminated by a unitary transform $U$, which is a diagonal matrix with the main diagonal defined as 
\begin{equation}
\label{eq:ZGNRProblemUnitaryTransform}
\{u_{2p -1,2p -1}, u_{2p, 2p}\}= \{(-1)^p,(-1)^p\}|_{p=1,\ldots, w} \, .
\end{equation}
For $w=2$, it reads
\begin{equation}
    U = \begin{pmatrix}
-1 & \phantom{-}0 & \phantom{-}0 & \phantom{-}0   \\
\phantom{-}0 & -1 & \phantom{-}0 & \phantom{-}0   \\
\phantom{-}0 & \phantom{-}0 & \phantom{-}1 & \phantom{-}0  \\
\phantom{-}0 & \phantom{-}0 & \phantom{-}0 & \phantom{-}1  
\end{pmatrix} \, .
\end{equation}
As follows from~\eqref{eq:ZGNRProblemUnitaryTransform}, $U$ is both a unitary and an involutory matrix. It can be straightforwardly checked that applying the unitary transform~\eqref{eq:ZGNRProblemUnitaryTransform} to the eigenvector of $H$ given by Eq.~\eqref{eq:ZGNRProblemHamiltonian}, i.e., $\left| \tilde{c}^{(j)} \right\rangle = U \left| c^{(j)} \right\rangle$, we obtain eigenvectors of the Hamiltonian $\tilde{H} = U H U^{\dagger}$. For $w=2$, the explicit form of the new Hamiltonian is
\begin{equation}
\tilde{H} = \label{eq:ZGNRProblemHamiltonian2}
\left(\begin{array}{cccccc}
0 & \gamma q & 0 & 0  \\
\gamma q & 0 & -\gamma & 0  \\
0 & -\gamma & 0 & \gamma q  \\
0 & 0 & \gamma q & 0 
\end{array}\right) \, .
\end{equation}
The general form of the eigenvectors of $\tilde{H}$ is the same as~\eqref{eq:ZGNRProblemEigenVectorComponents3} but without $(-1)^p$ factor:
\begin{align}
\label{eq:ZGNRProblemEigenVectorComponents4}
\tilde{c}^{(j)}_{2p-1} &= \mp (-1)^j \sin \left[ (w+1-p) \theta_j \right] \,; & &  \\
\nonumber
\tilde{c}^{(j)}_{2p} &= \sin p \theta_j  \, , \qquad p= 1,\dots, w \, .
\end{align}

Equations~\eqref{eq:ZGNRProblemEigenVectorComponents4} and~\eqref{eq:ZGNRProblemEigenVectorComponents3} present components of non-normalized eigenvectors $\left| c^{(j)} \right\rangle$. Normalization constant $N_j$ for these vectors can be found from the normalization condition $ N^2_j \left\langle c^{(j)} \right. \left| c^{(j)} \right\rangle = N^2_j \sum_{p=1}^w c^{(j) \ast }_{2 p -1} c^{(j)}_{2 p -1} + c^{(j) \ast }_{2 p} c^{(j)}_{2 p} =1$, which yields 
\begin{equation}
\label{eq:ZGNRProblemTBmodelNormalizationConstant}
N_j = \dfrac{1}{\sqrt{w-\cos \left[ (w + 1) \theta_j \right] \, \dfrac{\sin w \theta_j}{\sin \theta_j}}} \, .
\end{equation}
We do not use `` $\widetilde{\phantom{c}}$ '' two distinguish the two types of eigenvectors mentioned above because, by definition, unitary transform preserves the dot product, therefore the normalization constant is the same in both cases.

As in the case of the secular equation, eigenvectors and normalization constants for the missing solution $\theta$ are obtained by the substitution $\theta \rightarrow i \beta $, which results in wave functions being exponentially decaying from the ribbon edges to its interior. These wave functions describe the so-called edge states~\cite{Klein1994,Fujita1996,Nakada1996}. In contrast to them, the wave functions given by normal solutions $\theta_j$ extend over the whole ribbon width, therefore they describe the so-called extended or bulk states. It can be shown that normalized eigenvectors' components for extended and edge states seamlessly match in the transition point $k_t$ defined by $2 \cos (k/2) = w/(w+1)$ (see Appendix~\ref{app:sec:EigenVectorsAtTransitionPoint}). 

\begin{figure}[htbp]
      \includegraphics[width=0.47\textwidth]{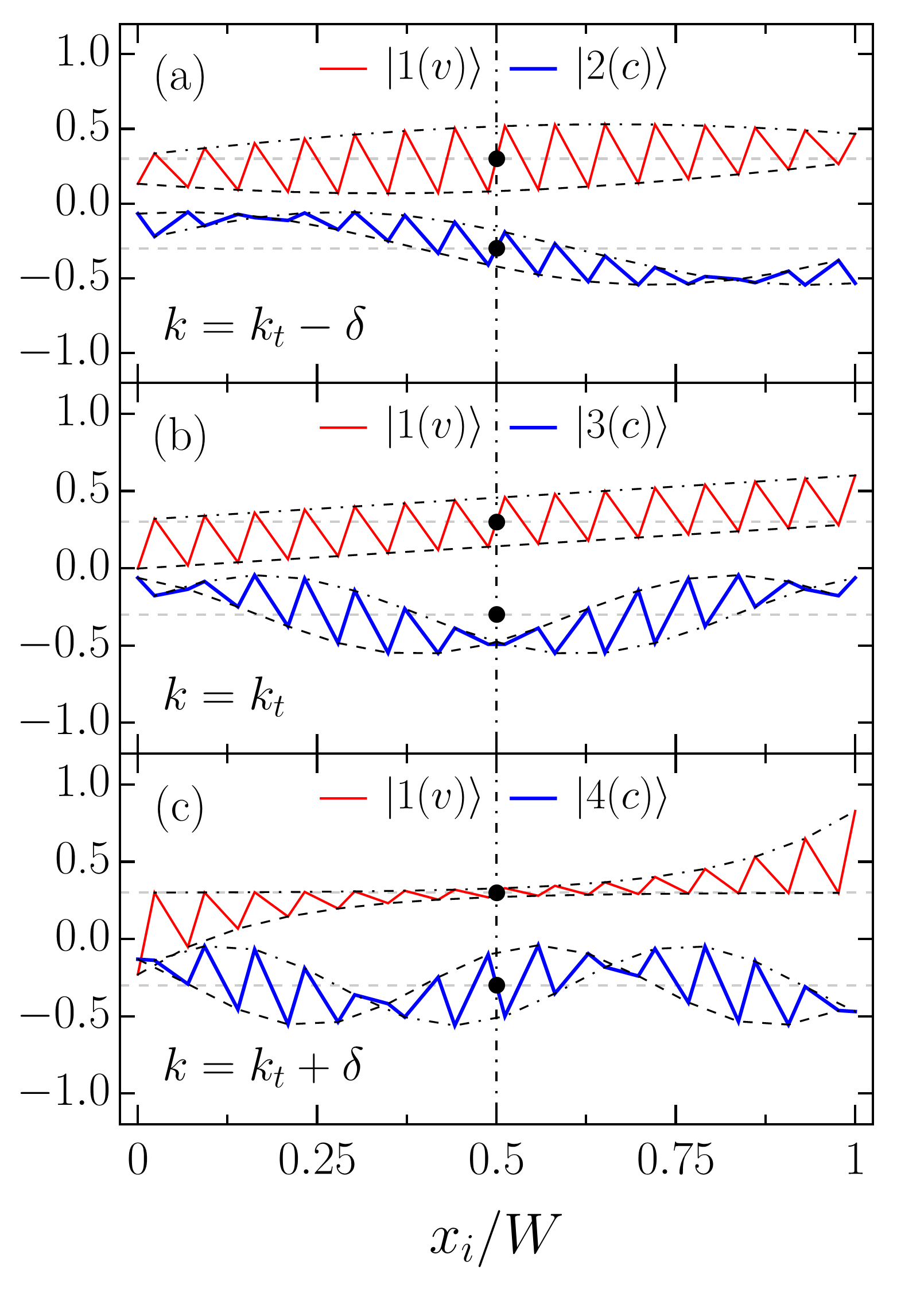} \\
	\caption{The bulk-edge transformation and parity of a zigzag nanoribbon wave function. The normalized wave functions $\left|J(s)\right\rangle$ of the zigzag nanoribbon with $w=15$ for various bands $J(s)$ and the Brillouin zone points $k=k_t + i \delta$: $\delta = 0.3$ (a) $i = -1$, (b)  $0$, and (c)  $1$. The solid lines are used for eye guidance, while the dashed and dashed-dotted curves represent the envelopes of the $2p-1$ (A) and $2p$ (B) sites. 
	The horizontal axis is a normalized transverse coordinate $x_i/W$, with $W$ being the ribbon width. The plots are shifted vertically by $\pm 0.3$ for clarity. The dashed dotted vertical line and thick black points denote the line of the mirror and centers of the inversion symmetry, respectively.
	}
	\label{fig:ZGNRWavefunctions}
\end{figure}

The matching of the bulk and edge state wave functions is shown in Fig.~\ref{fig:ZGNRWavefunctions}, where the wave functions of the zigzag ribbon with $w=15$ are plotted as functions of the atomic site positions $x_{2 p -1} = (\sqrt{3} a/2) (p-1)$ and $x_{2p} = (a/2\sqrt{3}) + x_{2p-1}$ normalized by the ribbon width $W = x_{2w}$. 
Figure~\ref{fig:ZGNRWavefunctions} presents wave functions for several energy branches $J(s)$, where $J$ is the energy branch number and $s=c$ or $v$ refers to the conduction or valence branch, respectively. As one can see, a bulk state wave function $\left|1(v)\right\rangle$, Fig.~\ref{fig:ZGNRWavefunctions} (a),  transforms into a wave function $\left| 1(v) \right\rangle$ predominantly concentrated at the ribbon edges and decaying towards the ribbon center, Fig.~\ref{fig:ZGNRWavefunctions} (c), by becoming a linear function of $x_i/W$ at $k=k_t$ as shown in Fig.~\ref{fig:ZGNRWavefunctions} (b). One can also see that the parity factor can be associated with the mirror or inversion symmetry of the electron wave function. For conduction subbands, if the parity factor $(-1)^J$ is positive, then the wave function is symmetric with respect to the inversion center denoted by the large black point as seen for $\left|2(c)\right\rangle$ and $\left|4(c)\right\rangle$ in Figs.~\ref{fig:ZGNRWavefunctions} (a) and~\ref{fig:ZGNRWavefunctions} (c). This means the wave function is odd. However, if $(-1)^J$ is negative, then the wave function is even, i.e., it is symmetric with respect to the reflection in the dashed dotted line signifying the ribbon center. This happens for $\left|3(c)\right\rangle$ in Fig.~\ref{fig:ZGNRWavefunctions} (b). For the valence subbands, the behavior is opposite: if $(-1)^J$ is negative then the state wave function is odd, as can be seen from Fig.~\ref{fig:ZGNRWavefunctions} for the subband $1(v)$, but it is even for positive parity factor $(-1)^J$. Such behavior is in agreement with the general properties of motion in one dimension~\cite{BookLandauVolIII1977}. The parity factor attributed to the mirror symmetry with respect to the line bisecting the ribbon longitudinally (see Fig.~\ref{fig:ZGNRAtomicStructureAndNumberingMFLin}) has been discussed in the literature~\cite{Kivelson1983,Hsu2007,Sanders2012,Sanders2013}. In this view, it should be noted that the unit cells of ribbons with odd $w$ do not have such a reflection symmetry (see Fig.~\ref{fig:ZGNRAtomicStructureAndNumberingMFLin} for $w=3$), nevertheless as we see from Fig.~\ref{fig:ZGNRWavefunctions}, for such ribbons, the wave functions can still be classified as even or odd in aforementioned sense. This suggests that the symmetry argument developed in Ref.~\onlinecite{Gundra2011a} as a criterion for the usage of the gradient approximation, which is to be discussed in the next section, is not complete, since in that form it applies only to ribbons with even $w$. Finally, we notice that the state wave functions can be classified by a number of twists of the envelope functions presented in Fig.~\ref{fig:ZGNRWavefunctions} by dashed and dashed dotted curves. The number of such twists (nodes) is equal to $J(c)$ and $J(v)-1$ for the conduction and valence subbands $J(s)$, respectively. This behavior is similar to what is expected from the oscillation theorem~\cite{BookLandauVolIII1977}.

\subsection{\label{subsec:OpticalMatrixElements}Optical transition matrix elements}
In this section, we study the optical properties of graphene nanoribbons with zigzag edges.  Optical transition matrix elements are worked out in the gradient (effective mass) approximation~\cite{Blount1962a,Johnson1973,LewYanVoon1993,Lin1994} and optical selection rules are obtained. However, before moving to the matrix elements of the ribbons, we shall introduce details of optical absorption spectra calculations where these matrix elements are to be used.

Within the first-order time-dependent perturbation theory the transition probability rate between two states, say $\left|\Psi_f \right. \rangle $ and $\left|\Psi_i \right. \rangle $ having energy $E_f$ and $E_i$, respectively, is given by the golden rule~\cite{BookAnselm1981}:
\begin{equation}
\label{eq:FermiGoldenRule}
A_{i\rightarrow f} = \dfrac{2 \pi}{\hbar} \left|\left\langle \Psi_f \left| \hat{H}_{\text{int}}\left(t\right) \right| \Psi_i \right\rangle \right|^2 \delta(E_f-E_i - \hbar \omega) \, ,
\end{equation}
where $\delta(\dots)$ is the Dirac deltafunction, and $\hat{H}_{\text{int}}\left(t\right)$ is a time-dependent interaction Hamiltonian coupling a system in question to that causing a perturbation, which is periodic in time with frequency $\omega$. Considering an incident plane electromagnetic wave as a perturbation, one can show in the dipole approximation, $e^{i \vec{k}\cdot \vec{r}} \approx 1$, that 
\begin{equation}
\left\langle \Psi_f \left| \hat{H}_{\text{int}}\left(t\right) \right| \Psi_i \right\rangle \sim \dfrac{E_0}{\omega} \left\langle \Psi_f \left| \hat{\vec{v}} \cdot \vec{e}_p \right| \Psi_i \right\rangle \equiv \dfrac{E_0}{\omega} M_{f,i} \, ,
\end{equation}
where $\hat{\vec{v}}$ is the velocity operator, $E_0$ is the electric field strength amplitude and $\vec{e}_p$ is the vector of electromagnetic wave polarization. Thus, optical transition matrix elements can be reduced to the velocity operator matrix elements (VMEs).

The total number of transitions per unit time in solids irradiated by electromagnetic wave at zero temperature is a sum of $A_{i\rightarrow f}$ over all initial (occupied) states in the valence band and final (unoccupied) states in the conduction band. To account for losses such as impurity and electron-phonon scattering, the deltafunction in Eq.~\eqref{eq:FermiGoldenRule} is replaced by a Lorentzian. The difference in occupation numbers of the initial and final states due to the finite temperature is introduced by the Fermi-Dirac distribution. Then, for the absorption coefficient due to the interband transitions, one has
\begin{widetext}
\begin{equation}
\label{eq:ZGNRProblemSpectralFunction}
A(\omega) \sim \sum_{n,m,k,s,s^{\prime}} \mbox{Im}\left[\dfrac{f(E_{m,s}(k))-f(E_{n,s^{\prime}}(k)) }{E_{n,s^{\prime}}(k) - E_{m,s}(k) - \omega - i \Gamma}\right] \dfrac{\left| M_{n(s),m(s^{\prime})} (k) \right|^2}{\omega} \, ,
\end{equation}
\end{widetext}
where $E_{m,s}(k)$ is the dispersion of the electron in the $m$-th conduction ($s=c$) or valence ($s=v$) subband, $f(E_{m,s}(k))$ is the Fermi-Dirac distribution function, $M_{n(s),m(s^{\prime})} (k)$ is the optical transition matrix element being a function of the electron wave vector, $\Gamma$ is the phenomenological broadening parameter ($0.004 \, \gamma $)~\cite{Lin2000}. Note that for nonzero temperature summation over initial states should also include states in the conduction band, therefore indices $s,s^{\prime}$ have been introduced above. The frequency of an incident wave, $\omega$, as well as the electron energy, is measured in the hoping integral $\gamma$.

Similar to Ref.~\cite{Chung2011}, we follow the prescription of the gradient approximation~\cite{Blount1962a,LewYanVoon1993} to obtain the velocity operator right from the system Hamiltonian:
\begin{equation}
\label{eq:OpticalMatrixElementsGradientApprox}
    \hat{\vec{v}} = \dfrac{i}{\hbar} [ \hat{H}, \hat{r}] =  \dfrac{1}{\hbar} \dfrac{\partial \hat{H}}{\partial \vec{k}} \, 
\end{equation}
whence for a one-dimensional case,
\begin{equation}
\label{eq:OpticalMatrixElementsGradientApprox1}
v = \dfrac{1}{\hbar} \dfrac{\partial H}{\partial k} \, ,
\end{equation}
with $H$ being the Hamiltonian of the unperturbed system. Note that the derivative $\partial H / \partial k$ is different from  $\partial H / \partial \vec{A}$ mentioned in Ref.~\cite{Sasaki2011}, where $\vec{A}$ is the vector potential. The former has a clear relation to the minimal coupling $\vec{k} \rightarrow \vec{k} + (e/\hbar) \vec{A}$ via the expansion $H(\vec{k}+(e/\hbar)\vec{A}) = H(\vec{k}) + (e/\hbar) \nabla_{\vec{k}} H \cdot \vec{A} + \ldots$, where higher-order terms can be neglected for small $\vec{A}$. Such an approach is equivalent to the effective mass treatment since the commutator $\left[\ldots \right]$ in Eq.~\eqref{eq:OpticalMatrixElementsGradientApprox} implies that the crystal momentum $\vec{k}$ is an operator:
\begin{equation}
\label{eq:ZGNRProblemCrystalMomentumOperator}
\vec{k} = \dfrac{1}{i}\dfrac{\partial \hphantom{p} }{\partial x} \vec{i} + \dfrac{1}{i}\dfrac{\partial \hphantom{p} }{\partial y} \vec{j} \, ,
\end{equation} 
which commutes with the position operator in the same way as real momentum $\vec{p}$, i.e. $\left[ x, k_x \right] = i$. Note, however, that there is no formal restriction to low energies around the Dirac point, $k=2 \pi /3$, as in the $\vec{k} \cdot \vec{p}$ theory with the effective mass approximation for graphene~\cite{DiVincenzo1984a,Ando2006a}, carbon nanotubes~\cite{Ajiki1993,Ajiki1996a}, or graphene nanoribbons~\cite{Brey2006,Palacios2010}. 

In what follows, we proceed with the calculation and analysis of the velocity operator matrix elements (VMEs) in the gradient (effective mass) approximation. Introducing the following vector:
\begin{equation}
\label{eq:ZGNRProblemZetaVector}
\left| \zeta^{(m)} \right\rangle =  \dfrac{a}{\hbar}\dfrac{\partial H (k)}{\partial k} \left| c^{(m)} \right\rangle \, ,
\end{equation}
the VME is evaluated as
\begin{align}
\label{eq:OpticalMatrixElementsVME}
M_{n(c),m(v)} &= \left\langle c^{(n)}_c \right. \left| \zeta^{(m)}_v \right\rangle \nonumber \\
 &= \sum_{p=1}^w \underset{c}{c^{(n) \ast }_{2 p -1}} \underset{v}{\zeta^{(m)}_{2 p -1}} + \underset{c}{c^{(n) \ast }_{2 p}} \underset{v}{\zeta^{(m)}_{2 p}} \, ,
\end{align}
where indices ``$c$'' and ``$v$'' denote the conduction and valence band, respectively, and the eigenvectors $\left| c^{(n,m)} \right\rangle $ are meant to be normalized. In Eq.~\eqref{eq:ZGNRProblemZetaVector}, the graphene lattice constant $a$ emerged because, in contrast to the general expression~\eqref{eq:OpticalMatrixElementsGradientApprox1}, the electron wave vector $k$ is now treated again as a dimensionless quantity.

Let us calculate VMEs for the Hamiltonian $\tilde{H}$ of the form presented by Eq.~\eqref{eq:ZGNRProblemHamiltonian2}. Similar calculations for $H$ results in the same final expression. Due to the nature of unitary transforms it is not essential which of the Hamiltonians and corresponding eigenvectors one uses. The components of vectors $\left| \tilde{\zeta}^{(j)} \right\rangle$ are
\begin{align}
\label{eq:ZGNRProblemZetaVectorComponentsMFLinHamiltonian}
\tilde{\zeta}^{(j)}_{2p-1} &=  - \dfrac{\gamma a}{\hbar} \sin (\dfrac{k}{2}) \sin p \theta_j \,, \qquad p=1,\dots, w \, ;   \nonumber \\
\tilde{\zeta}^{(j)}_{2p} &= \pm \dfrac{\gamma a}{\hbar} \sin (\dfrac{k}{2}) (-1)^j \sin \left[  (w+1-p) \theta_j \right] \, ,  
\end{align}
with upper ``$+$'' ( lower ``$-$'') being used for conduction (valence) subbands. Substituting Eqs.~\eqref{eq:ZGNRProblemEigenVectorComponents4} and~\eqref{eq:ZGNRProblemZetaVectorComponentsMFLinHamiltonian} into~\eqref{eq:OpticalMatrixElementsVME}, one obtains
\begin{align}
\label{eq:VMETBmodel}
M_{n(c),m(v)} &= \dfrac{\gamma a}{\hbar} \sin (\dfrac{k}{2}) N_n N_m \left[ (-1)^n - (-1)^m \right] S_{n,m}  \, ,
\end{align}
where $S_{n,m}$ is a sum. A similar form of the matrix element was obtained in Ref.~\onlinecite{Chung2011} but explicit expressions for the sum $S_{n,m}$ and normalization constants $N_{n}(N_{m})$ were not provided and potential singularities in VME due to $N_j$ and $S_{n,m}$ dependence on $k$ were not analysed. Such an analysis has not been carried out elsewhere including Ref.~\onlinecite{Sasaki2011}.

It is known that the topological singularity in the graphene wave functions~\cite{Ando2002,Ando2005} leads to anisotropic optical matrix element and absorption in the vicinity of the Dirac point~\cite{Gruneis2003,Saito2004}. This anisotropy is eliminated in the matrix element of carbon nanotubes~\cite{Gruneis2003,Jiang2004}, but the matrix element can exhibit singular behavior at the Dirac point of the tube's Brillouin zone if a perturbation such as strain, curvature~\cite{Portnoi2015} or external magnetic field~\cite{Portnoi2009,Moradian2010,Chegel2014} is applied. The sharp dependence of the zigzag ribbon VME on the electron wave vector around $k= \pm 2 \pi/3$ could be triggered by the presence of the edge states. This possibility, however, has not been analysed yet. The VME behavior at the transition point $k_t$ has not been investigated either. Being of practical interest~\cite{Portnoi2015} this requires a thorough analysis of possible singularities in the VME dependence on $k$. The $S_{n,m}$ sum is given by
\begin{align}
\label{eq:OpticalMatrixElementsSum}
S_{n,m} &= \sum_{p=1}^w \sin \left[ (w+1-p) \theta_n \right] \sin p \theta_m \, ; \\ \nonumber
&=\dfrac{\sin \theta_m \sin \left[ (w+1) \theta_n \right] - \sin \left[ (w+1) \theta_m \right] \sin \theta_n}{2 (\cos \theta_n - \cos \theta_m)} \, .
\end{align}
In Eq.~\eqref{eq:VMETBmodel}, normalization constants have been added since the vectors given by Eq.~\eqref{eq:ZGNRProblemEigenVectorComponents4} and used for obtaining Eq.~\eqref{eq:ZGNRProblemZetaVectorComponentsMFLinHamiltonian} are not normalized. It is important to allow for normalization constants in the VMEs because otherwise due to their $\theta_j$ and therefore $k$ dependency the VME curve's behavior in the vicinity of the transition point $k_t$ is incorrect. It is also worth noting that for $\theta_n = \theta_m$, or equivalently for $S_{n,n}$, there is an indeterminacy of $\tfrac{0}{0}$ type in the summation result of Eq.~\eqref{eq:OpticalMatrixElementsSum}. This indeterminacy can be easily resolved by L'Hospital's rule, which yields
\begin{equation}
S_{n,n} = \dfrac{(w+2) \sin w \theta_n - w \sin \left[ (w+2) \theta_n \right]}{4 \sin \theta_n} \,.
\end{equation}
In a similar fashion, one can check that for $\theta_n \rightarrow 0$, $S_{n,n} \rightarrow 0$. Note, however, that if $\theta_n \rightarrow 0$, then the normalization constant $N_n$ given by Eq.~\eqref{eq:ZGNRProblemTBmodelNormalizationConstant} becomes infinitely large, thereby introducing indeterminacy into the VME. For transitions between the valence and conduction subbands this indeterminacy is not essential for it is multiplied by an exact zero, originating from the square brackets in Eq.~\eqref{eq:VMETBmodel}, which ensures a zero final result.

As can be seen from Eq.~\eqref{eq:VMETBmodel}, $M_{n(c),n(v)}$ is zero, whereas $M_{n(c),n+1(v)} \sim N_n N_m S_{n,m} \sin (k/2)$. Thus, optical selection rules are: if $\Delta J = n-m$ is an even integer, then transitions are forbidden, whereas if $\Delta J = n-m$ is an odd integer, then transitions are allowed. The influence of the factor $S_{n,m}$ together with the normalization constants $N_n$ and $N_m$ on the transition probability, omitted in Ref.~\onlinecite{Chung2011}, will be discussed in detail in Sec.~\ref{sec:Discussion}. In the remainder of this section, we consider transitions between only conduction (valence) subbands, which are considered in Ref.~\onlinecite{Sasaki2011} but are beyond the scope of Ref.~\onlinecite{Chung2011}.

If the temperature is not zero, then there is a nonzero probability to find an electron in the conduction subband states. Therefore, an incident photon can be absorbed due to transitions between conduction subbands. The same is true for valence subbands, which are not fully occupied. That is why, as has been pointed out above, the summation in Eq.~\eqref{eq:ZGNRProblemSpectralFunction} is to be carried out over transitions between conduction (valence) subbands too. Thus, for the absorption coefficient calculation, one also needs VMEs for such transitions. Making use of Eqs.~\eqref{eq:ZGNRProblemEigenVectorComponents4} and~\eqref{eq:ZGNRProblemZetaVectorComponentsMFLinHamiltonian}, we obtain
\begin{align}
\label{eq:ZGNRProblemVMEvvcc}
M_{n(s),m(s)} &=\left\langle c^{(n)}_s \right. \left| \zeta^{(m)}_s \right\rangle \, ; \\
 &= \pm \dfrac{\gamma a}{\hbar} \sin (\dfrac{k}{2}) N_n N_m \left[ (-1)^n + (-1)^m \right] S_{n,m}  \, , \nonumber
\end{align}
where ``$+$'' and ``$-$'' are used for VME of transitions between conduction, $s=c$, and valence, $s=v$, subbands. For the specified transitions, the optical selection rules are the following: transitions are allowed if $\Delta J$ is an even number and they are forbidden otherwise. These matrix elements and corresponding selection rules should be important in spontaneous emission (photoluminescence) calculations~\cite{Oyama2006}.

In the case of $n=m$, VME given by Eq.~\eqref{eq:ZGNRProblemVMEvvcc} is nothing else but the group velocity of an electron in the $n$-th band. If $n=m=1$, then $\theta_n = \theta_m \rightarrow 0$ as $k$ approaches the transition point $k_t$. As a result, in Eq.~\eqref{eq:ZGNRProblemVMEvvcc}, the indeterminacy arises in precisely the same manner as discussed above for Eq.~\eqref{eq:VMETBmodel}. In the present case, however, it is essential since the expression in square brackets of Eq.~\eqref{eq:ZGNRProblemVMEvvcc} is not an exact zero. The indeterminacy can be resolved by the application of L'Hospital's rule twice. This burden, however, can be bypassed by calculating the VME by the aid of simplified expressions for eigenvectors at the $k_t$ provided in Appendix~\ref{app:sec:EigenVectorsAtTransitionPoint}. Such a calculation yields
\begin{equation}
\label{eq:ZGNRProblemGroupVelocity}
M_{1(s),1(s)} = \mp \dfrac{\gamma a}{\hbar} \sin (\dfrac{k_t}{2})  \dfrac{w+2}{2 w + 1} \, ,
\end{equation}
where the upper (lower) sign is used for the conduction (valence) subband. It is easily seen from the expression above that in the limit of a wide ribbon the electron group velocity at $k_t \approx 2 \pi / 3$, i.e., approaching to the Dirac point,  is $\mp v_F/2$.

Velocity matrix elements for transitions involving edge states can be easily obtained from Eqs.~\eqref{eq:VMETBmodel} and~\eqref{eq:ZGNRProblemVMEvvcc} with $S_{n,m}$ given by Eq.~\eqref{eq:OpticalMatrixElementsSum} after  $\theta \rightarrow i \beta$ replacement being applied. It should be noticed that the Eqs.~\eqref{eq:VMETBmodel} and~\eqref{eq:ZGNRProblemVMEvvcc} obtained here are incomparably simpler than their analogues in Ref.~\cite{Sasaki2011} [cf. with Eqs.~(18) and~(19) therein]. In the next section, we discuss and investigate numerically the obtained results.

\section{\label{sec:Discussion}Numerical Results and Discussion}
\subsection{Electronic properties}
\begin{figure}[htbp]
	\includegraphics[scale=0.35]{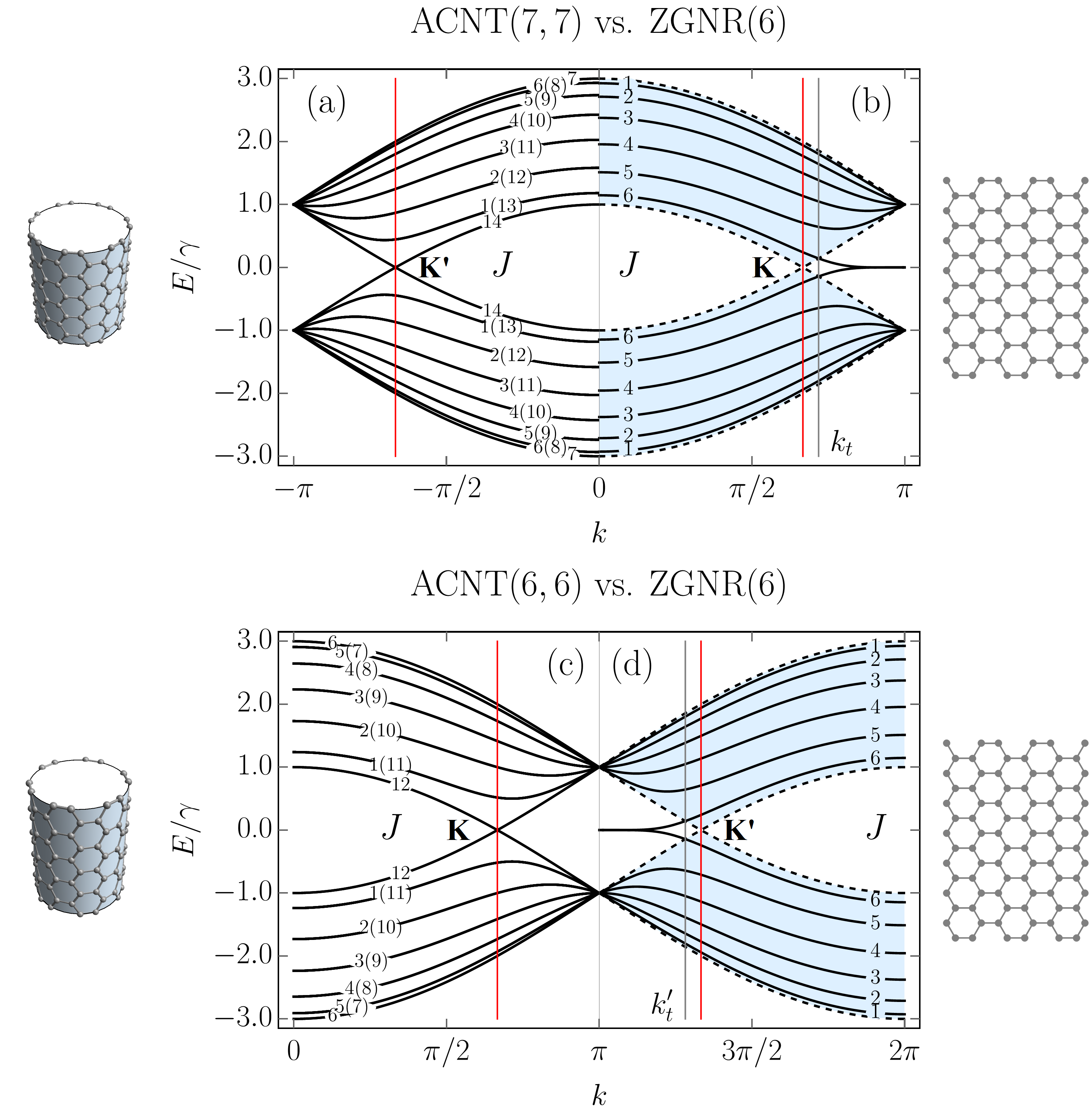}
	\caption{\label{fig:TubeVsRibbonBandStructure} A zigzag nanoribbon and armchair nanotube band structure matching. (a) The band structure of an armchair carbon nanotube, ACNT$(7,7)$, compared to (b) that of a zigzag ribbon with $w=6$, ZGNR$(6)$. (c) and (d) The same as (a) and (b) but for ACNT$(6,6)$. The dashed gray curves encompass light blue area, which signifies the region of the graphene band structure. The vertical lines $k_t$ and $k^{\prime}_t$ mark positions of the transitions points defined by equation $2 \cos (k/2) = w/(w+1)$ in the vicinity of {\bf K} and {\bf K}$^{\prime}$ points (i.e., $k=\pm 2 \pi /3$), respectively. The inverse band numbering for the ribbon used in Appendix~\ref{app:sec:PeriodicBoundaryConditions} and direct band numbering for the tube, i.e., for $A=-\cos \theta$, are shown. The corresponding atomic structures are presented on both sides for clarity.}
\end{figure}

The physical properties of graphene nanoribbons are often related to those of carbon nanotubes (CNTs). In particular, one usually compares the electronic properties of graphene nanoribbons with those of carbon nanotubes~\cite{Hsu2007,Sasaki2009}. In most cases, such a comparison is based merely on the fact that an unrolled carbon tube transforms into a graphene ribbon. However, this approach is a crude one. Firstly, because only zigzag (armchair) ribbons with even number of carbon atom pairs can be rolled up into armchair (zigzag) tubes. Secondly, because a more relevant and subtle comparison requires the matching of boundary conditions. It has been shown by White \emph{et al.}~\cite{White2007a} that periodic and ``hard wall" boundary conditions can be matched for armchair ribbons and zigzag carbon nanotubes if the width of the ribbons is approximately equal to half of the circumference of the tubes. In Fig.~\ref{fig:TubeVsRibbonBandStructure}, we demonstrate that a similar correspondence of the electronic properties takes place for zigzag graphene nanoribbons with $w$ zigzag chains, ZGNR($w$), and armchair carbon nanotubes, ACNT$(w+1,w+1)$ and ACNT$(w,w)$ depending on which parts of the Brillouin zones are matched (see Appendix~\ref{app:sec:PeriodicBoundaryConditions}). The impossibility of matching a zigzag ribbon with just one of the tubes arises from the secular equation~\eqref{eq:ZGNRProblemQuantizationCondition} linking transverse wave vector $\theta$ with the longitudinal wave vector $k$.
For sure, due to the presence of the edge states, one should not expect the transport properties of undoped ribbons to be the same as those of tubes, but the equivalence of the optical properties seems to be quite natural thing. However, this is not the case. As was shown numerically~\cite{Lin2000,Hsu2007,Chung2011,Chung2016} and has been demonstrated above analytically, the optical selection rules of zigzag ribbons are different from those of armchair tubes~\cite{Ajiki1994,Vukovic2002,Gruneis2003,Jiang2004,Malic2006} (see also Appendix~\ref{app:sec:ArmchairCNTselectionRules}). This leads to transitions between the edge states being forbidden, which should also have important implications for zigzag ribbon based superlattices~\cite{Saroka2016a,Saroka2015,Saroka2014a}. A somewhat similar picture is observed in the bilayer graphene quantum dots of triangular shape, where the edge states are dispersed in energy around the Fermi level~\cite{Abdelsalam2016}.

\begin{figure}[htbp]
    \includegraphics[scale=0.4]{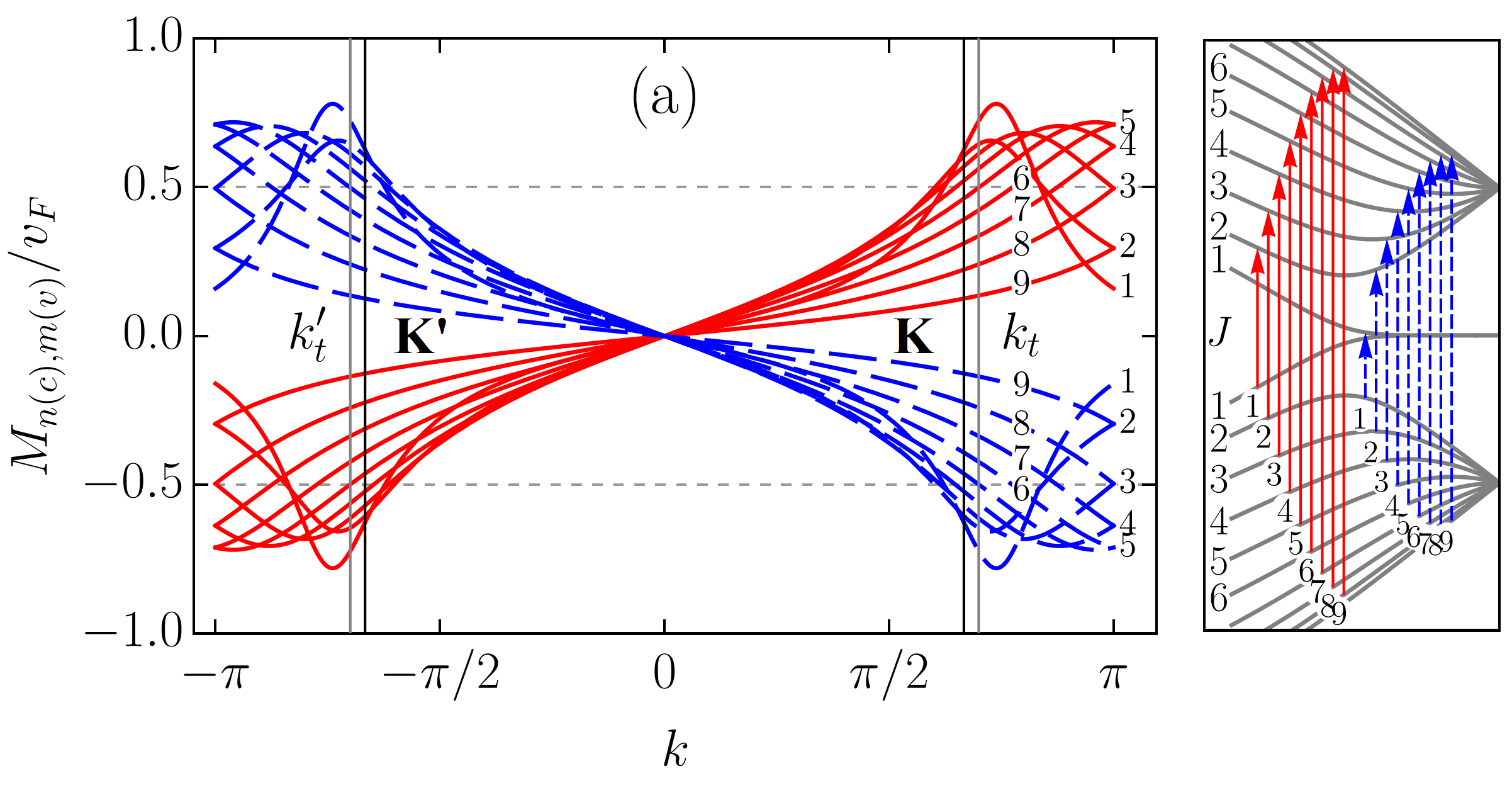} \\
    \includegraphics[scale=0.4]{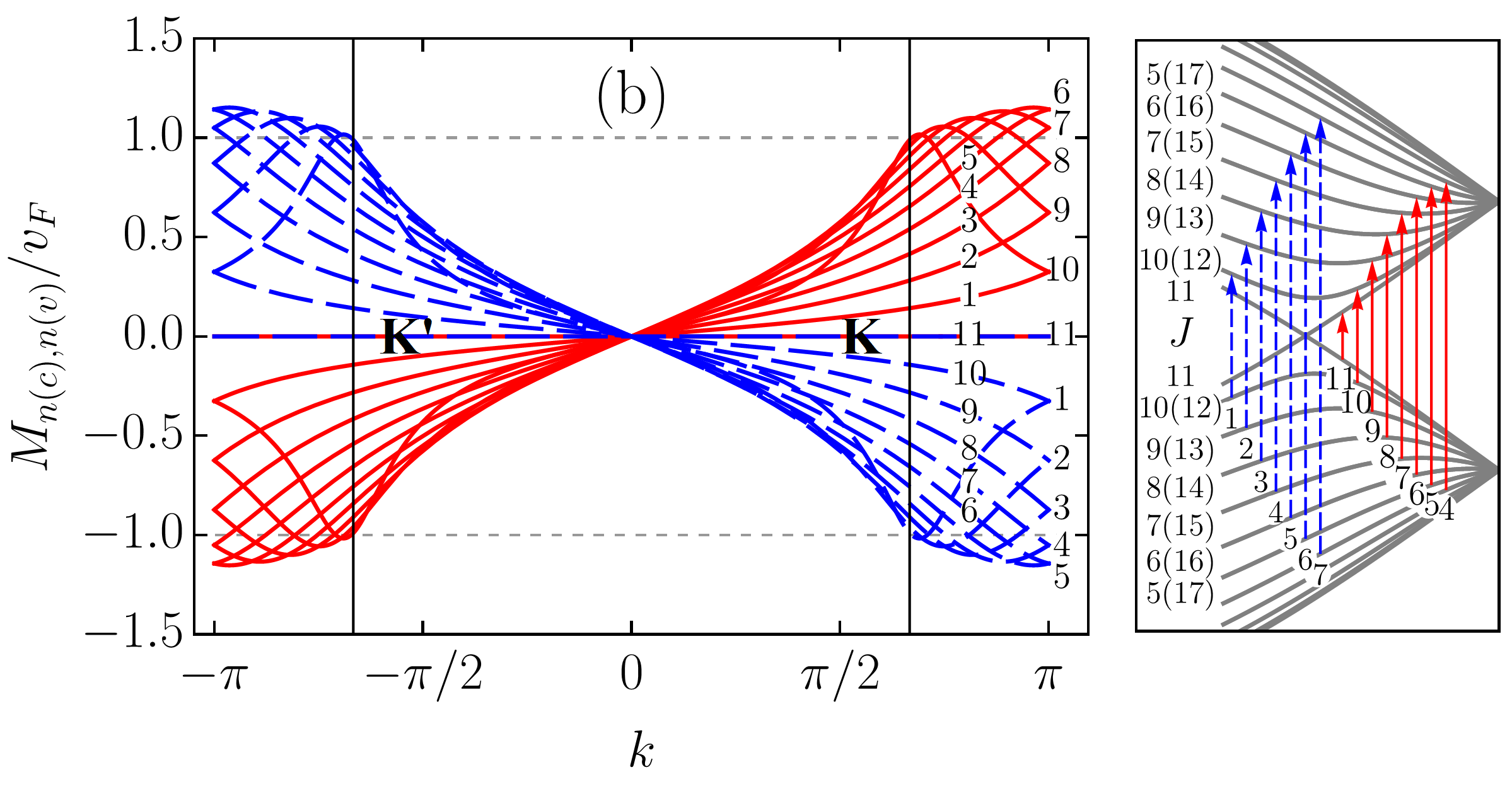} 
	\caption{\label{fig:ZGNRProblemVMEsInTBmodel} The velocity operator matrix elements of a zigzag nanoribbon and armchair carbon nanotube with similar $k$ dependence. (a) The VMEs of ZGNR$(10)$ transitions $v \rightarrow c; \Delta J = 1$  within the first Brillouin zone in comparison with (b) those of ACNT$(11,11)$ transitions $v \rightarrow c; \Delta J = 0$. The labels of the VME curves correspond to those of vertical arrows presenting the transitions in the right panels. The index $J$ shows the direct band numbering resulting from Eq.~\eqref{eq:ZGNRProblemCosTheta} for the ribbon and inverse numbering for the tube (see Appendix~\ref{app:sec:PeriodicBoundaryConditions}). The double degenerate tube's bands have two labels. Dashed arrows represent transitions between the bands numbered in round brackets.
	}
\end{figure}

\begin{figure}[htbp]
    \includegraphics[scale=0.4]{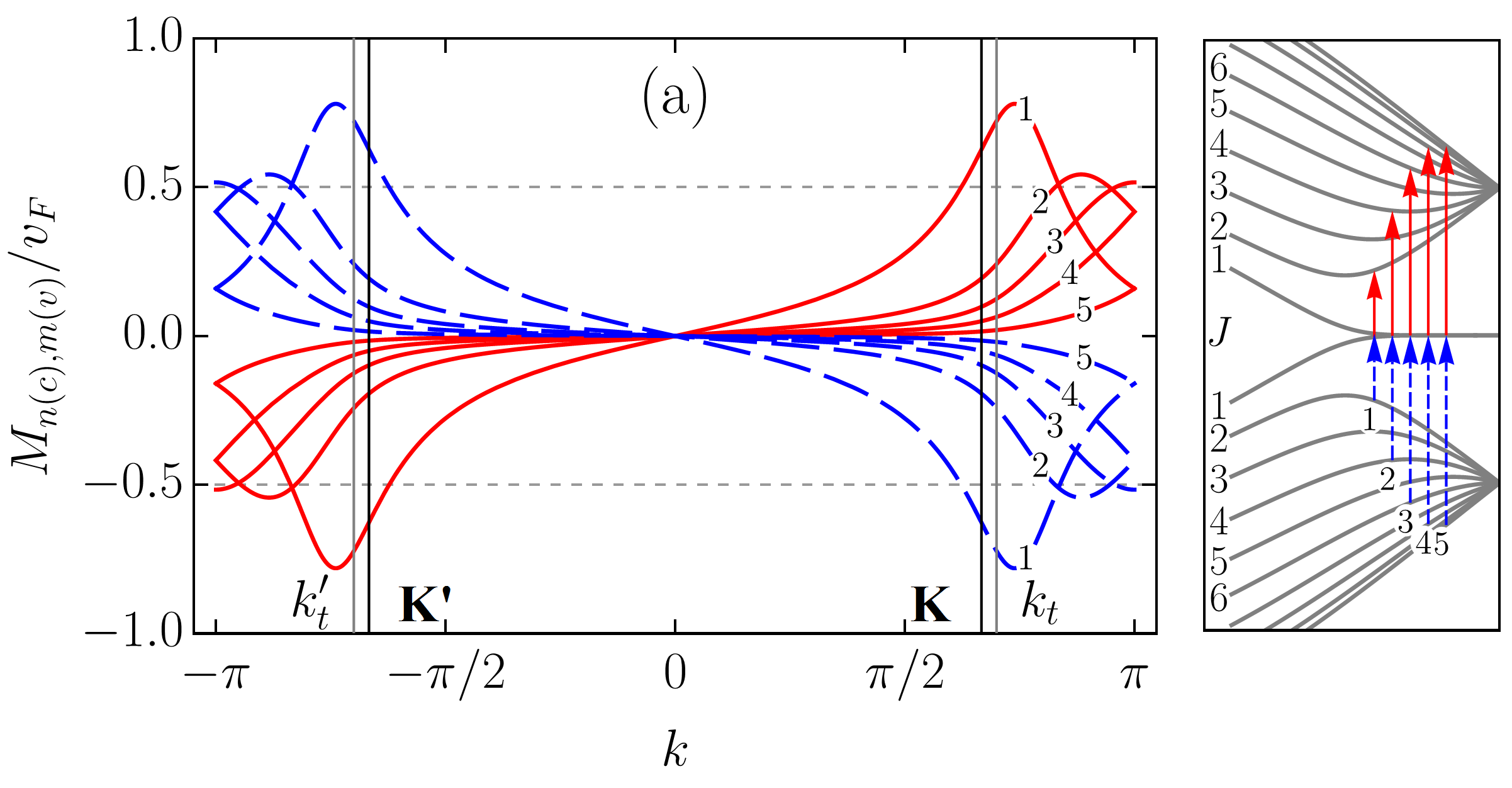} \\
    \includegraphics[scale=0.4]{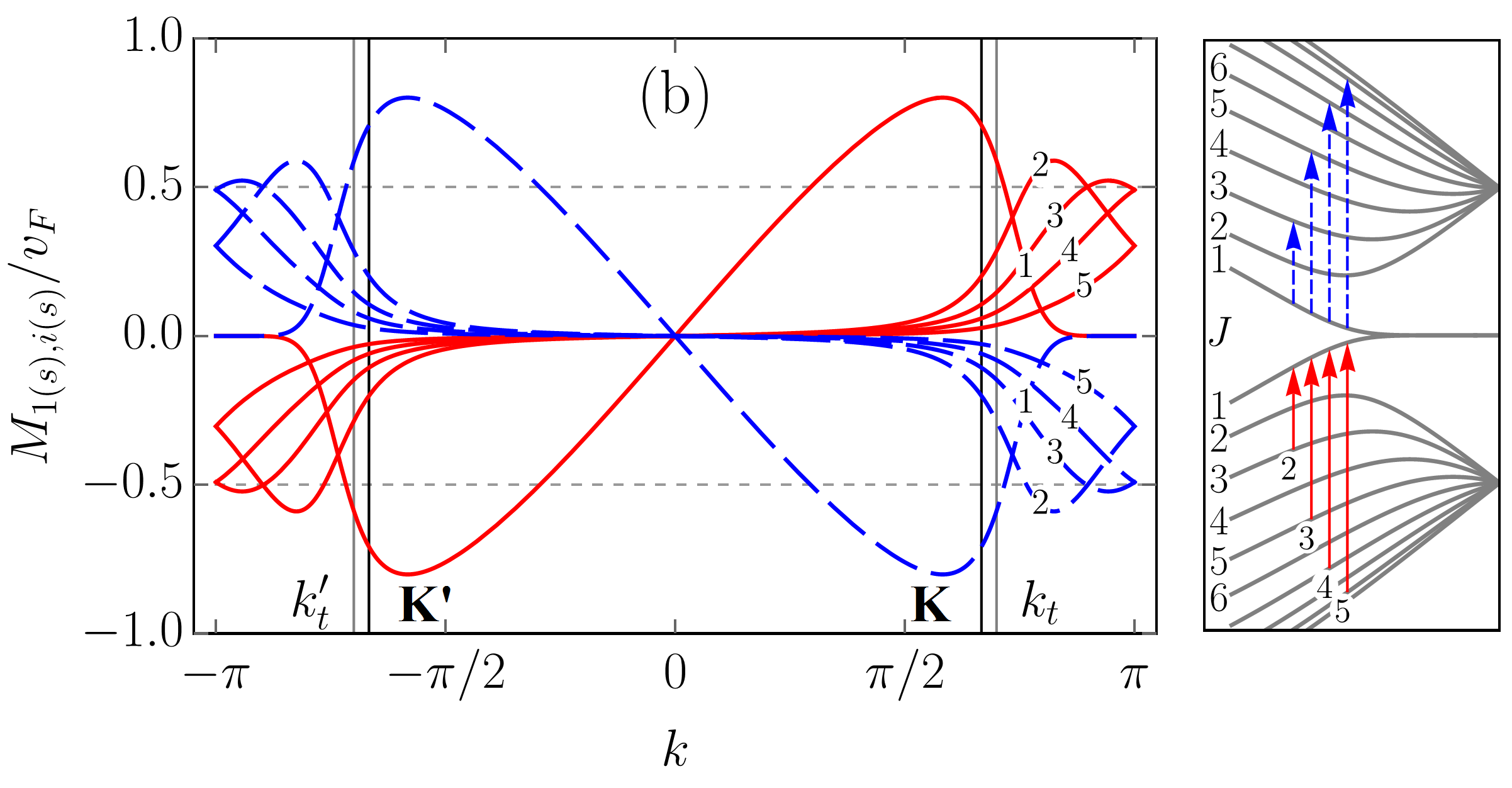} \\
	\caption{\label{fig:ZGNRProblemVMEsInTBmodel2}The velocity operator matrix elements for transitions inherent to zigzag ribbons. The VMEs of the allowed transitions of ZGNR$(10)$ within the first Brillouin zone: (a) $v \rightarrow c; \Delta J = 1,3,5,\ldots$. (b) $v \rightarrow v; c \rightarrow c; \Delta J = 0,2,4, \ldots$. The VME curves and energy band labeling follows the same convention as in Fig.~\ref{fig:ZGNRProblemVMEsInTBmodel}.
	}
\end{figure}

\subsection{Optical properties}
\subsubsection{Optical transition matrix elements}
To scrutinize the velocity operator matrix elements (VMEs) for allowed transitions we focus on the zigzag ribbon with $w=10$. In Figs.~\ref{fig:ZGNRProblemVMEsInTBmodel} and~\ref{fig:ZGNRProblemVMEsInTBmodel2}, we plotted the VMEs given by Eqs.~\eqref{eq:VMETBmodel} and~\eqref{eq:ZGNRProblemVMEvvcc} as functions of the electron wave vector in the first Brillouin zone (BZ). Figure~\ref{fig:ZGNRProblemVMEsInTBmodel} includes results for an armchair tube for the sake of comparison. All plots are normalized by the graphene Fermi velocity $v_F = \sqrt{3} a \gamma / (2 \hbar)$. The arbitrary phase factor of the VMEs, which does not affect their absolute values, was chosen such that it favours plots' clarity. As in previous sections, we follow the adopted two index notation for the ribbon bands: $J(s)$, where $J = 1,\ldots, w$ is the band number and $s=c,v$ is the band type with ``$c$" and ``$v$" standing for conduction and valence bands, respectively. With this notation in mind one can see that the VME curves for transitions $j(v) \rightarrow (j+1)(c)$ [$(j+1)(v) \rightarrow j(c)$], where $j = 1, \dots, w-1$ are shown in Fig.~\ref{fig:ZGNRProblemVMEsInTBmodel} (a). The VME curves for transitions $1(v) \rightarrow 2 n(c)$ [$2 n(v) \rightarrow 1(c)$], where $n = 1,\dots, w/2 \text{ or } (w-1)/2$, and for transitions between conduction (valence) subbands only, i.e., $1(s) \rightarrow (2 n - 1)(s)$, where $n = 1, \dots, w/2 \text{ or } (w-1)/2$ are presented in Figs.~\ref{fig:ZGNRProblemVMEsInTBmodel2} (a) and~\ref{fig:ZGNRProblemVMEsInTBmodel2} (b), respectively. As one can see, the VME curves deviate significantly from the previously reported  $\sin (k/2)$ behavior~\cite{Chung2011,Sasaki2011}, according to which extrema are to be at $k=\pi$, i.e., at the edge of the BZ. The deviation is due to the $S_{n,m}$ and $N_j$ given by Eqs.~\eqref{eq:OpticalMatrixElementsSum} and~\eqref{eq:ZGNRProblemTBmodelNormalizationConstant} [see also Eq.~\eqref{eq:ZGNRProblemTBmodelNormalizationConstant3}], respectively. The shift of the VME curve extrema from the BZ edge is larger for low-energy transitions. Interestingly enough, the positions of these extrema in BZ do not coincide with those of the energy band extrema resulting in the van Hove singularities in the density of states. The curves labeled by $1$ in Figs.~\ref{fig:ZGNRProblemVMEsInTBmodel} (a) and~\ref{fig:ZGNRProblemVMEsInTBmodel2} (a) represent direct transitions from the edge states to the closest in energy bulk states. These curves have the largest magnitudes among the ribbons VMEs. However, even for them the maximum absolute values are well below $v_F$, in sharp contrast to what is seen in Fig.~\ref{fig:ZGNRProblemVMEsInTBmodel} (b) for ACNT$(11,11)$ (cf. with Refs.~\cite{Jiang2004,Portnoi2009}). Though it is difficult to ignore the fact that the shapes of the VME curves $2$ to $9$ in Fig.~\ref{fig:ZGNRProblemVMEsInTBmodel} (a) are very similar to those obtained for ACNT VMEs in Fig.~\ref{fig:ZGNRProblemVMEsInTBmodel} (b). The most profound curves in Fig.~\ref{fig:ZGNRProblemVMEsInTBmodel2} (b) are also labeled by $1$, but they do not have corresponding transitions depicted in the panel to the right. This is because these curves are, in fact, the electron group velocities in $1(v)$ and $1(c)$ subbands given by Eq.~\eqref{eq:ZGNRProblemVMEvvcc}. As can be seen, at the transition points $k_t$ and $k_t^{\prime}$ marked by vertical lines the group velocity curves have magnitudes about $v_F/2$. This is in accordance with Eq.~\eqref{eq:ZGNRProblemGroupVelocity}. Ignoring the group velocity curve, one finds that the most prominent magnitudes of VME have transition $1(c)\rightarrow 3(c)$ [$3(v) \rightarrow 1(v)$]. The probability rate described by VMEs of $1(s) \rightarrow (2n-1) (s)$, where $n=2,\dots$, transitions is comparable to that of transitions $1(v) \rightarrow 2 n(c)$ [$1(c) \rightarrow 2 n(v)$], where $n=2,\dots$, labeled by $2$, $3$, $4$, etc., in Fig.~\ref{fig:ZGNRProblemVMEsInTBmodel2} (a) and~\ref{fig:ZGNRProblemVMEsInTBmodel2} (b). However, these transitions are less intense compared to $1(v) \rightarrow 2(c)$ [$2(v) \rightarrow 1(c)$], or majority of the $j(v) \rightarrow (j+1)(c)$ [$(j+1)(v)  \rightarrow  j(c)$], where $j = 1, \dots, w-1$, transitions presented in Fig.~\ref{fig:ZGNRProblemVMEsInTBmodel} (a). A regular smooth behavior of all matrix elements at the {\bf K}({\bf K}$^\prime$) and $k_t$ ($k_t^{\prime}$) points is worth highlighting, especially for those including $1(s)$ subbands. We noticed, however, that for increasing ribbon width (up to $w=25$) the VME curve peaks for transitions involving $1(s)$ subbands gain a sharper form, therefore a singular VME behavior may still be expected for $1(v) \rightarrow 2(c)$ [$2(v) \rightarrow 1(c)$] transitions in ribbons with $w>25$.

\subsubsection{Absorption}
It follows from Figs.~\ref{fig:ZGNRProblemVMEsInTBmodel} and~\ref{fig:ZGNRProblemVMEsInTBmodel2} (see also Appendix~\ref{app:sec:SuplementaryResults})  that the absorption spectra of zigzag ribbons are mostly shaped by $v \rightarrow c$ transitions with $\Delta J=1$ presented in Fig.~\ref{fig:ZGNRProblemVMEsInTBmodel} (a). However, other transitions may play an important role at certain conditions created by interplay of the doping (or temperature) and ribbon width. To check this, we investigated optical absorption spectra given by Eq.~\eqref{eq:ZGNRProblemSpectralFunction} for narrow ribbons with $w = 2 \ldots10$. In what follows we discuss ZGNR($6$) for it has the most prominent features and additionally it has been recently synthesized with atomically smooth edges~\cite{Ruffieux2016}.

\begin{figure*}[tbp]
      \includegraphics[scale=0.64]{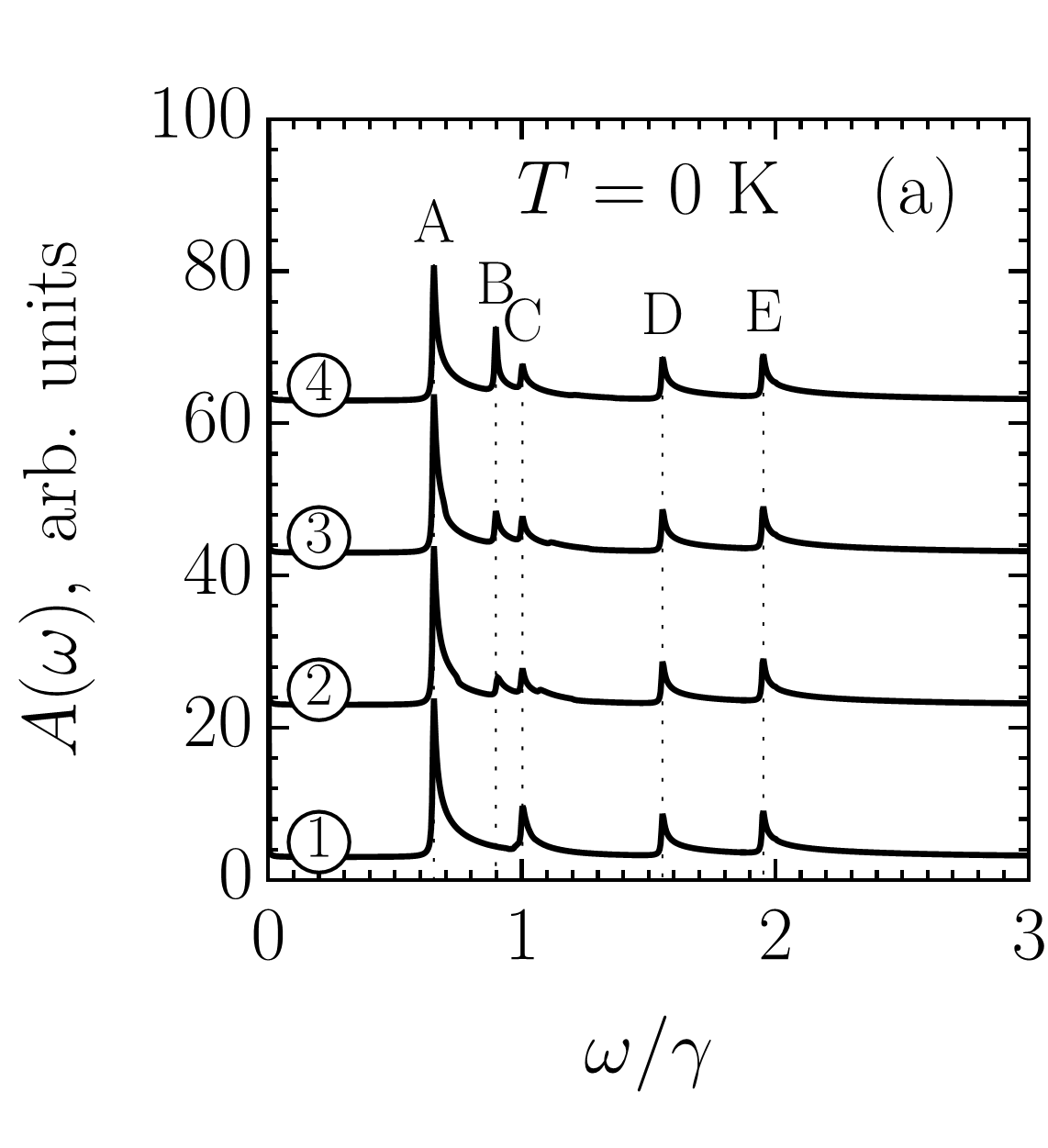}
      \includegraphics[scale=0.64]{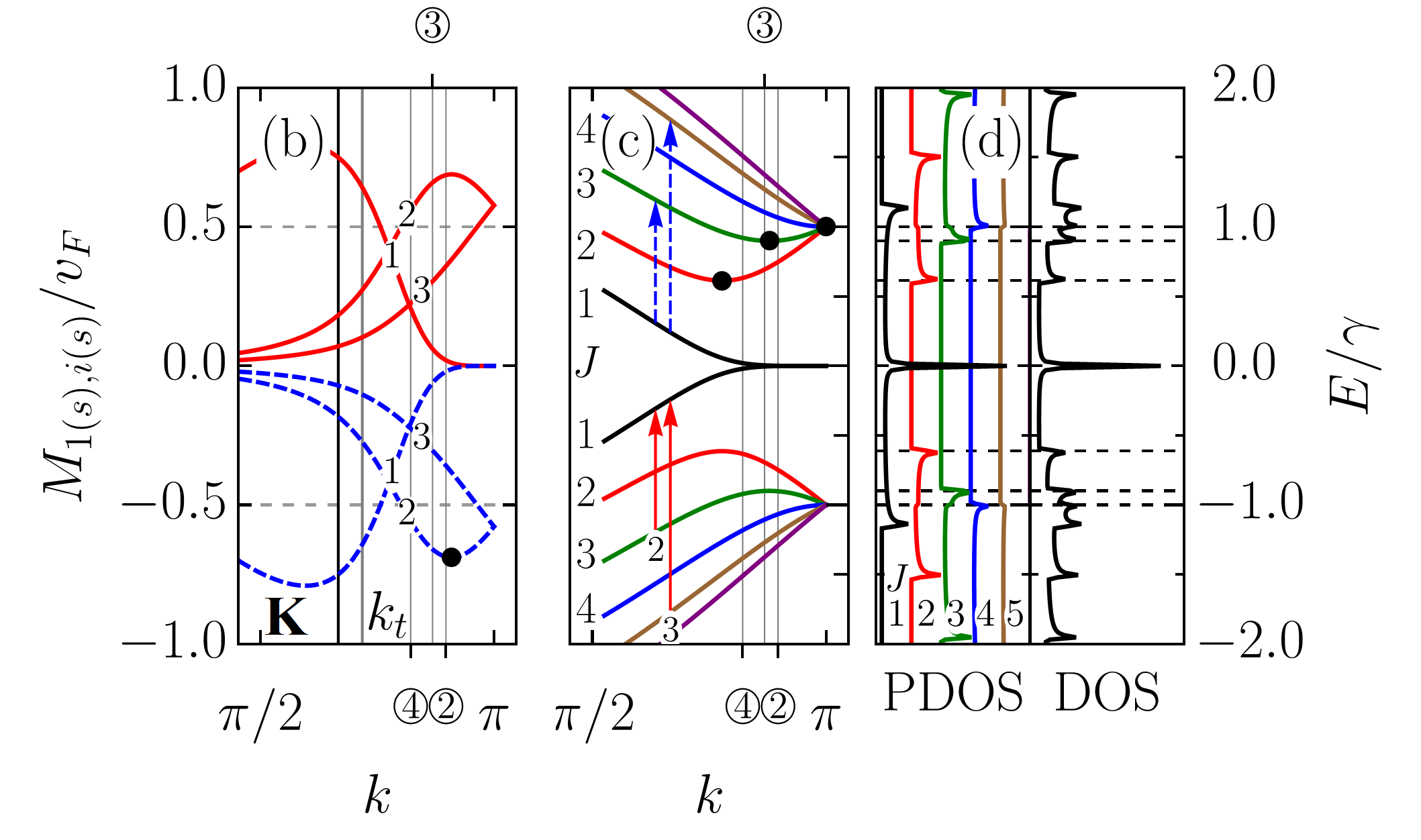}
	\caption{The doping-dependent absorption peaks in zigzag graphene nanoribbons. (a) The absorption spectra of ZGNR($6$) for various positions of the Fermi level: $E_{F} = 0$, $0.001 \gamma$, $0.004 \gamma$, and $0.02 \gamma$ for the curves \protect\textcircled{\raisebox{-1.2pt}{1}}, \protect\textcircled{\raisebox{-1.2pt}{2}}, \protect\textcircled{\raisebox{-1.2pt}{3}}, and \protect\textcircled{\raisebox{-1.2pt}{4}}, respectively. The frequency $\omega$ is measured in hopping integrals $\gamma$. The spectra are shifted vertically for clarity. (b) The VMEs for transitions depicted in (c) the band structure of ZGNR($6$). The vertical lines labeled by encircled numbers mark the positions of the points where the Fermi levels cross the $1(c)$ subband. The thick black points signify subband and VME extrema. (d) The partial, i.e., for each subband separately, and total density of states for ZGNR($6$). The color and number of the partial density of states curves correspond to those of the relevant subbands presented in (c); these curves are also offset horizontally for clarity.}
	\label{fig:ZGNRProblemSpectralFunctionFLdependence}
\end{figure*}
Figure~\ref{fig:ZGNRProblemSpectralFunctionFLdependence} compares the absorption spectra of ZGNR($6$) for various positions of the Fermi level, $E_F$. As one can see, depending on $E_F$ the absorption spectrum has $4$ or $5$ pronounced peaks, which we label in ascending order of their frequency as $A$, $B$, $C$, $D$, and $E$. Peaks $D$ and $E$ are not sensitive to the doping, whereas peaks $A$, $B$, and $C$ are. In contrast to peaks $A$ and $C$ undergoing suppression with increasing $E_F$, peak $B$ significantly strengthens. Such different behavior of the three peaks is explained by their different nature. 

Let us start with the most interesting case of the peak $B$ at $\omega= 0.9 \gamma$, which corresponds to the wavelength of about $400$~nm if $\gamma \approx 3 $~eV. This peak stems from transitions $1(c)\rightarrow3(c)$.  At $T=0$~K, the valence subbands are  fully occupied, therefore, we can safely exclude from the consideration transition $3(v) \rightarrow 1(v)$, which must be blocked due to the exclusion principle. The steep doping dependence of the peak $B$ observed in Fig.~\ref{fig:ZGNRProblemSpectralFunctionFLdependence} (a) has two causes. Firstly, dispersion of subbands $1(c)$ and $3(c)$ and resulting density of states $\sim \left(\partial E_{j,s} (k)/ \partial k\right)^{-1}$ presented in Fig.~\ref{fig:ZGNRProblemSpectralFunctionFLdependence} (d). Secondly, the nonzero VMEs for transition $1(c) \rightarrow 3(c)$ in the $k$-interval $\left(2 \pi /3, \pi \right)$, as shown in Fig.~\ref{fig:ZGNRProblemSpectralFunctionFLdependence} (b). 

Without doping the peak $B$ is absent in the absorption spectrum because both subbands $1(c)$ and $3(c)$ are empty. The introduction of doping results in large number of edge states in the almost flat subband $1(c)$ being occupied with electrons. If the point of the Fermi level intersection with the subband $1(c)$ is denoted as $k_F$, then one can say that $k_F$  rapidly shifts towards the {\bf K} point upon ribbon doping. In Figs.~\ref{fig:ZGNRProblemSpectralFunctionFLdependence} (b) and~\ref{fig:ZGNRProblemSpectralFunctionFLdependence} (c), the values of $k_F$ for $E_F = 0.001\gamma$, $0.004\gamma$, and $0.02 \gamma$ are marked by vertical lines labeled as \textcircled{\raisebox{-1.2pt}{2}}, \textcircled{\raisebox{-1.2pt}{3}}, and \textcircled{\raisebox{-1.2pt}{4}}, correspondingly. 
As seen in Fig.~\ref{fig:ZGNRProblemSpectralFunctionFLdependence} (b) at $E_F = 0.001\gamma$, i.e., $k_F=$\textcircled{\raisebox{-1.2pt}{2}}, VME of $1(c) \rightarrow 3(c)$ transition represented by curve `$2$' is close to the maximum magnitude, nevertheless, the intensity of the peak $B$ in Fig.~\ref{fig:ZGNRProblemSpectralFunctionFLdependence} (a) presented by curve \textcircled{\raisebox{-1.2pt}{2}} is not that large. The low intensity at such a level of doping is related to the fact that the subband $3(c)$ has a dispersion to the right of the vertical line \textcircled{\raisebox{-1.2pt}{2}}, which leads to transitions although being strong contribute into absorption at different frequencies. Upon further increase of the $E_F$ up to $0.02\gamma$, i.e., $k_F=$\textcircled{\raisebox{-1.2pt}{4}}, the VME for $1(c) \rightarrow 3(c)$ transition decreases in magnitude to about $v_F/2$. However, due to the flatness of subband $3(c)$ in the vicinity of the band minimum [thick black point in Fig.~\ref{fig:ZGNRProblemSpectralFunctionFLdependence} (c)], all the transitions  between lines \textcircled{\raisebox{-1.2pt}{2}} and \textcircled{\raisebox{-1.2pt}{4}} contribute into absorption nearly at the same frequency, which corresponds to the van Hove singularity in the density of states shown in Fig.~\ref{fig:ZGNRProblemSpectralFunctionFLdependence} (d). This results in the sharp enhancement of the peak $B$.

The filling of the subband $1(c)$ with electrons affects all the transitions: $1(c) \rightarrow 3(c)$, $5(c)$, etc. However, in ZGNR($6$) the higher order transition $1(c)\rightarrow 5(c)$ is buried in the peak $C$ for it has lower density of states compared to the subband $4(c)$. To observe higher order transitions one has to take a wider ribbon. Any of the ribbons $w=8,9,10$ can be chosen but ribbon with $w=9$ is the best choice for there transitions $1(c)\rightarrow 5(c)$ results in a clear peak at $\omega \approx \gamma$.

According to our calculations, ZGNR($6$) and ZGNR($7$) are the best choices for a detection of the tunable peak due to $1(c) \rightarrow 3(c)$ transitions. The latter is in agreement with the results of Sanders \emph{et al.}~\cite{Sanders2012,Sanders2013} based on the matrix elements of the momentum and with the wave function overlapping taken into account. For wider ribbons, the peak broadens and loses intensity due to combined effect of the VME and density of states reduction.

As for peaks $A$ and $C$ at $\omega= 0.65\gamma$ and $\gamma$ in Fig.~\ref{fig:ZGNRProblemSpectralFunctionFLdependence} (a), they arise from interband transitions $1(v)\rightarrow2(c)$ [$2(v)\rightarrow1(c)$] and $1(v)\rightarrow4(c)$ [$4(v)\rightarrow1(c)$], respectively. Strictly speaking, many subbands converge into $E= \pm \gamma$ at $k=\pi$, therefore some other transitions also contribute into the peak $C$. By mentioning only one type of transition, we mean the dominant contribution in terms of density of states as indicated in Fig.~\ref{fig:ZGNRProblemSpectralFunctionFLdependence} (d). The intensity of the peak $C$ decreases with doping for it results in the subband $1(c)$ being filled with the electrons whereby transitions  $4(v) \rightarrow 1(c)$ are blocked due to the exclusion principle. The same Pauli blocking also takes place for transitions  $2(v) \rightarrow 1(c)$, therefore, intensity of the peak $A$ decreases too. A more gentle decrease of peak $A$ intensity compared to that of peak $C$ is due to low doping. As one can see in Fig.~\ref{fig:ZGNRProblemSpectralFunctionFLdependence} (c),  for the chosen values of the Fermi level the point $k_F$ does not reach position of the subband $2(c)$ minimum. For larger doping, $A$-peak intensity decreases as it happens for peak $C$, and it totally disappears if the doping is high enough to attain the $2(c)$ subband. 

The effect of the finite temperature is similar to that of doping discussed above (see Appendix~\ref{app:sec:SuplementaryResults}).

\begin{figure}[htbp]
      \includegraphics[width=0.47\textwidth]{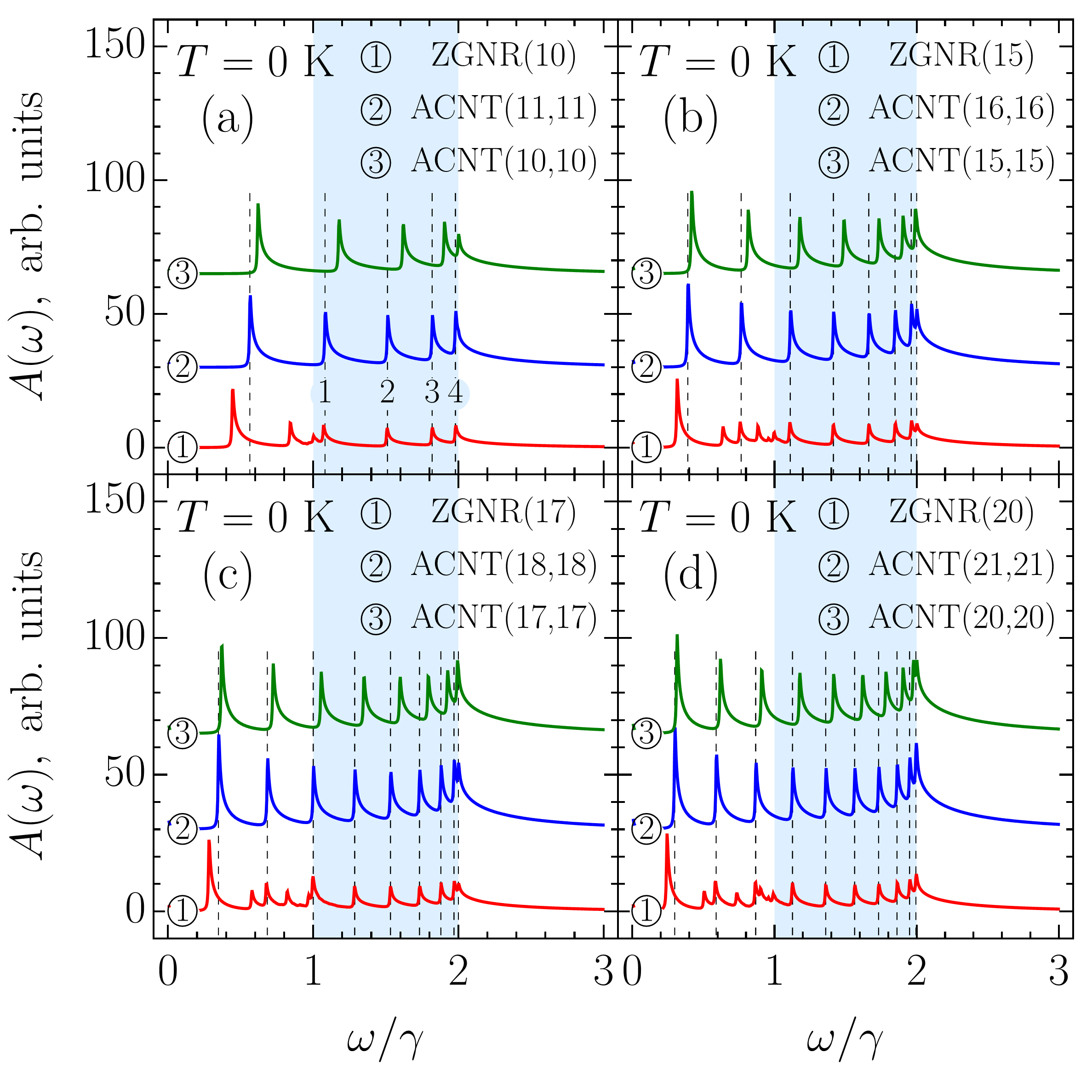} \\
      \includegraphics[width=0.47\textwidth]{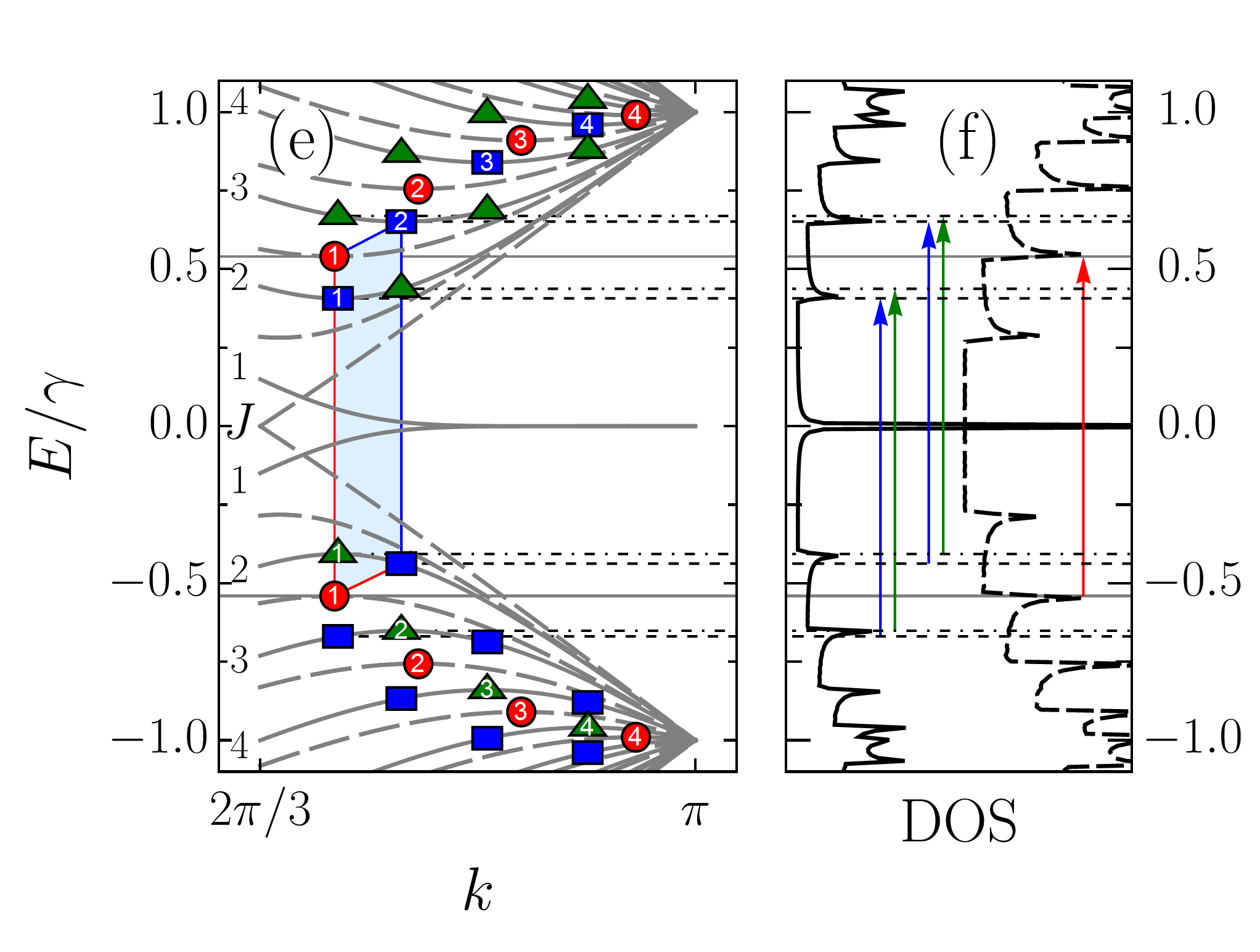}
	\caption{ The absorption peak correlation in zigzag nanoribbons and armchair nanotubes. (a)--(d) The absorption spectra of ZGNR($w$) compared to those of ACNT$(w+1,w+1)$ and ACNT$(w,w)$ for various ribbon widths and $E_F=0$. Absorption spectra are shifted vertically for clarity. (e) and (f) The band structure and the density of states for ZGNR$(10)$ (solid) and ACNT$(11,11)$ (dashed). The density of states curves are offset for clarity. The numbered circles denote the positions of the van Hove singularities in the tube. The numbered squares and triangles denote the van Hove singularities in the conduction and valence subbands of the ribbon, respectively. Transitions $v \rightarrow c$ are possible only between the markers of the same shape.}
	\label{fig:ZGNRvsACNTSpectralFunction}
\end{figure}

Finally, let us compare the zigzag nanoribbon absorption spectra with those of armchair nanotubes. In Fig.~\ref{fig:ZGNRvsACNTSpectralFunction} (a), the absorption spectra of ZGNR$(10)$ and ACNT$(11,11)$ are presented together with that of ACNT$(10,10)$. For the sake of comparison, each spectrum is not normalized by the number of atoms in the unit cell. The first peculiarity, which one can notice, is that in the ribbon all but the lowest in energy absorption peaks lose approximately half of their intensity compared to the peaks in the tubes. The second peculiarity is that ZGNR$(10)$ and ACNT$(11,11)$ have the same pattern of absorption peaks in the high frequency range $\omega > \gamma$, which is highlighted in the light blue. Both features are not accidental, as follows from the plots presented in Fig.~\ref{fig:ZGNRvsACNTSpectralFunction} (b)-\ref{fig:ZGNRvsACNTSpectralFunction}(d) for ribbons and tubes of larger transverse size.

In order to explain the noticed difference and similarity, we focus on ZGNR$(10)$ and ACNT$(11,11)$. Obviously, a large difference in peak intensities between the tube and ribbon cannot be explained only by the velocity matrix elements being higher in the tube than in the ribbon, as follows from Fig.~\ref{fig:ZGNRProblemVMEsInTBmodel}, therefore the density of states should be accounted for. Here we do not appeal to the suppression due to the momentum conservation as in Ref.~\onlinecite{Sasaki2011} for we regard all transitions, even between subbands with different indices, as direct ones. At the same time, the correlation of the absorption peaks' positions is to be related to the van Hove singularities in the density of states too. Thus we need to have a closer look at the band structures and density of states of ZGNR$(10)$ and ACNT$(11,11)$. In Fig.~\ref{fig:ZGNRvsACNTSpectralFunction} (e), the ZGNR$(10)$ band structure (solid curve) is compared with that of ACNT$(11,11)$ (dashed curve). Similar comparison is presented for the density of states in Fig.~\ref{fig:ZGNRvsACNTSpectralFunction}(f). The peaks numbered as $1$, $2$, $3$, and $4$ in Fig.~\ref{fig:ZGNRvsACNTSpectralFunction} (a) result from the transitions between ACNT$(11,11)$ subband extrema marked by numbered circles in Fig.~\ref{fig:ZGNRvsACNTSpectralFunction} (e). The same peaks in ZGNR$(10)$ originate from the transitions involving the subband extrema marked in Fig.~\ref{fig:ZGNRvsACNTSpectralFunction} (e) by the numbered squares (triangles) for the conduction (valence) subbands. Selection rules in both structures allow transitions between the markers of the same shape. Let us be more specific and focus on the peak `$1$'. In ACNT$(11,11)$, this peak is due to transition between two van Hove singularities in the density of states. Although the density of states in the tube is nearly twice as high as than that in the ribbon due to the double degeneracy of the tube's subbands, this cannot explain the difference in the intensities of the tube and ribbons absorption peaks, since, according to the selection rules, two type of transitions with the same frequency are allowed in the ribbon: $3(v)\rightarrow 2(c) [2(v)\rightarrow 3(c)]$. The difference in intensities arises due the fact that positions of the band extrema for adjacent bands in the ribbon are shifted in the $k$-space, thereby each of the specified in Fig.~\ref{fig:ZGNRvsACNTSpectralFunction} (e) ribbon transitions happens either `from' or `to' the band extrema and not `between' them as happens in the tube. In other words, each of these transitions involves only one van Hove singularity. In this view, the extremely high intensity of the lowest in energy absorption peak in ZGNR$(10)$ arises due to the high density of states originating from the flatness of the $1(c)$ band dispersion at $E=0$.

As one can notice from Fig.~\ref{fig:ZGNRvsACNTSpectralFunction} (e), the tube subband extrema take middle positions in energy between extrema of adjacent ribbons subbands. This leads to the tube's and ribbon's transition energies being very close as illustrated by a parallelogram in Fig.~\ref{fig:ZGNRvsACNTSpectralFunction} (e). As a result, a correlation between the absorption peak positions arises. To understand the origin of this correlation we need to analyze the positions of the van Hove singularities, which can be derived from the analytical expression for the band structure. However, for a zigzag ribbon, such an expression cannot be obtained in a closed-form from Eq.~\eqref{eq:ZGNRProblemProperEnergy}, since the secular equation~\eqref{eq:ZGNRProblemQuantizationCondition} does not allow expressing of its solution in such a form; though closed-form solutions for two specific cases, $k=2 \pi /3$ and $\pi$, have been reported for this type of equation~\cite{Klein1994,Compernolle2003,Wakabayashi2012}. On the other hand, since the armchair nanotube band structure has a closed-form given by Eq.~\eqref{app:ACNTBandStructure}, the positions of the van Hove singularities and, therefore, the absorption peak positions can be easily obtained for them. Then, a simple analytical expression for ACNT$(w+1,w+1)$ peak positions,
\begin{equation}
\label{eq:ZGNRProblemAbsorptionPeakPositions}
\widetilde{\omega}_{j} = 2 \gamma \sin [\pi j/(w+1)] \, ,
\end{equation}
can be used as an estimation of the absorption peak positions in ZGNR($w$), when $5/6>j/(w+1)>1/6$. In Fig.~\ref{fig:ZGNRvsACNTSpectralFunction} (a), the vertical dashed lines denote the peak positions given by Eq.~\eqref{eq:ZGNRProblemAbsorptionPeakPositions}. As one can see, outside the light blue regions peak positions do not necessarily coincide; the ribbon spectra also have additional peaks outside these regions resulting from transitions involving $1(s)$ subbands and the selection rule $v \rightarrow c$ $\Delta J=1,3,\ldots$, etc. In contrast to this, within the regions $\gamma < \omega < 2 \gamma$ the above-mentioned correlation takes place for all ribbons with $w > 5$. To estimate the reliability of Eq.~\eqref{eq:ZGNRProblemAbsorptionPeakPositions} in Table~\ref{tab:ZGNRProblemPeakPositions} we compared the numerically calculated peaks positions in the ZGNR$(10)$ with those resulting from Eq.~\eqref{eq:ZGNRProblemAbsorptionPeakPositions}. We supplemented these results with numerically evaluated energies of $j(v) \rightarrow (j + 1)(c)$, where $j=2,3,4,5$, transitions involving one band extremum state, i.e., those which occur between the states denoted by square ($\Box$) and triangle ($\triangle$) markers in Fig.~\ref{fig:ZGNRvsACNTSpectralFunction} (e). As seen from Table~\ref{tab:ZGNRProblemPeakPositions}, a deviation of $\widetilde{\omega}_j$ from $\omega_i$ does not exceed $1\%$ of the hopping integral, i.e., $30$~meV for $\gamma \approx 3$~eV. It also follows from the Table~\ref{tab:ZGNRProblemPeakPositions} that the above presented picture is a simplified one. In reality, the absorption peaks are averages of all transitions taking place in between the two subband extrema shifted in the $k$-space so that peak positions $\omega_i$ and their estimates $\widetilde{\omega}_j$ are squeezed between the $j(v) \rightarrow (j + 1)(c); (\Box, \triangle)$ transition energies and the energy differences between the corresponding van Hove singularities, $\Box_{i}-\triangle_{i+1}$.
\begin{table}
\caption{ \label{tab:ZGNRProblemPeakPositions} The absorption peak positions of ZGNR$(10)$ in the region $\omega > \gamma$ compared to the estimate $\widetilde{\omega}_j$ given by Eq.~\eqref{eq:ZGNRProblemAbsorptionPeakPositions} and transition energies between the states $j(v)\rightarrow (j+1)(c)$ denoted by $\Box$'s and $\triangle$'s in Fig.~\ref{fig:ZGNRvsACNTSpectralFunction} (e). The index $i$ numbers the peaks in Fig.~\ref{fig:ZGNRvsACNTSpectralFunction} (a). The last column presents the energy differences between the numbered subband extrema in Fig.~\ref{fig:ZGNRvsACNTSpectralFunction} (a). All quantities are measured in the hopping integral $\gamma$.}
\begin{ruledtabular}
\begin{tabular}{ccccccc}
$i$ & $j$ & $\omega_i$  & $\widetilde{\omega}_{j}$ & $\Box_v \rightarrow \Box_{c}$ & $\triangle_{v} \rightarrow \triangle_{c}$ & $\Box_{i}-\triangle_{i+1}$ \\
\hline
 1 & 2 &$1.074$ & $1.081$ & $1.089$ & $1.076$ & $1.058$ \\
 2 & 3 &$1.509$ & $1.511$ & $1.527$ & $1.518$ & $1.491$ \\
 3 & 4 &$1.821$ & $1.819$ & $1.839$ & $1.833$ & $1.799$ \\
 4 & 5 &$1.983$ & $1.980$ & $2.000$ & $1.998$ & $1.959$ \\
\end{tabular}
\end{ruledtabular}
\end{table}

\begin{figure}[htbp]
      \includegraphics[width=0.47\textwidth]{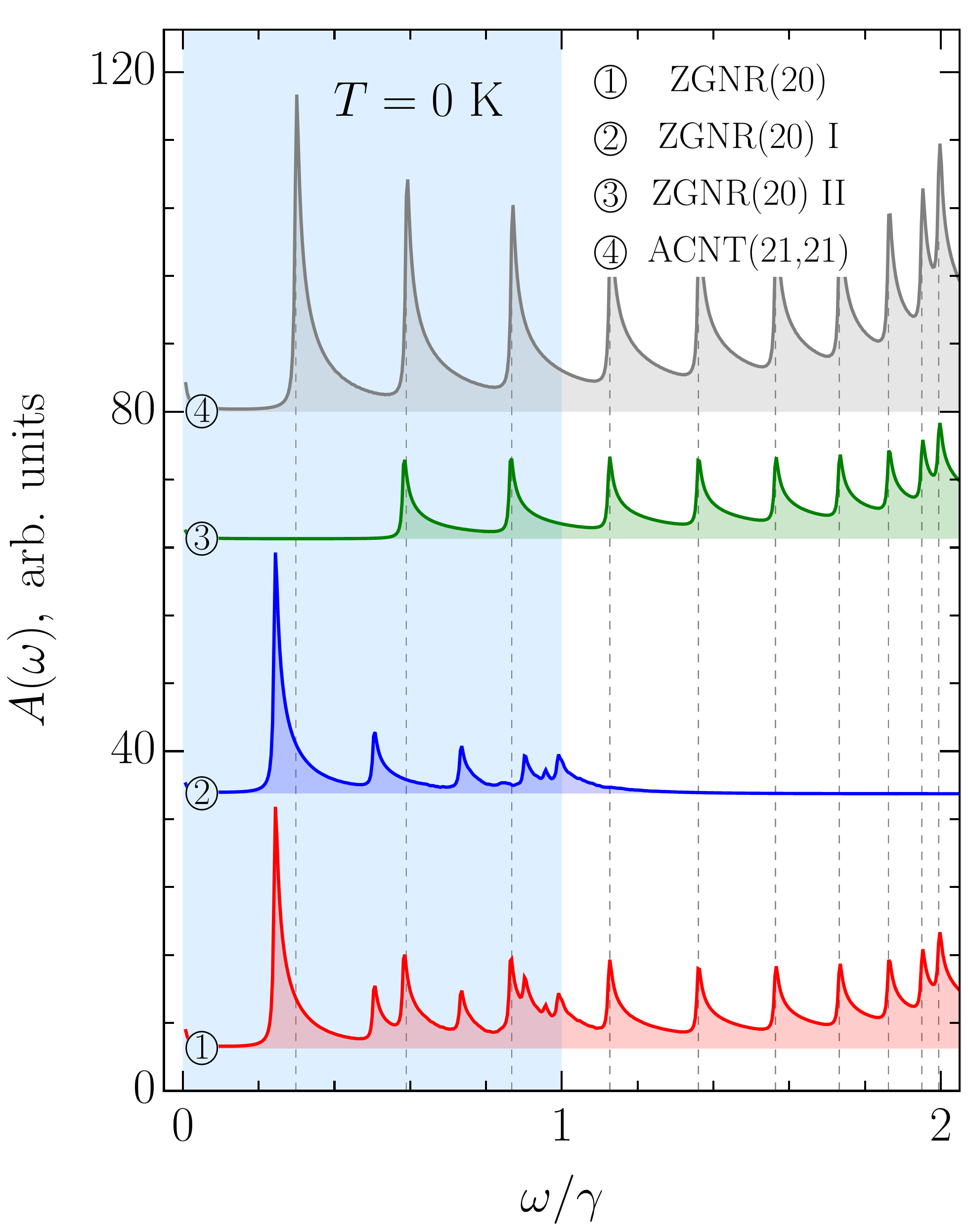} 
	\caption{ The low-energy absorption peak correlation in zigzag nanoribbons and armchair nanotubes. Absorption spectra are shifted vertically for clarity. The roman numbers (I) and (II) label spectra with only the edge states contribution and the part without it. The light blue region signifies the low-energy region where the correlation is hidden by the edge states transitions.}
	\label{fig:ZGNRvsACNTwithEdgeStateAbsorptionSeparated}
\end{figure}

The panels (b)-(d) in Fig.~\ref{fig:ZGNRvsACNTSpectralFunction} show that the aforementioned correlation may extend to the low-energy region $\omega < \gamma$. This region of a ribbon's spectrum is dominated by the transitions originating from the edge states. It is evident that the absorption peaks originating from these transitions cannot correlate with the peaks in armchair tubes. In fact, they can only hide this feature. In order to verify our assumption, in Fig.~\ref{fig:ZGNRvsACNTwithEdgeStateAbsorptionSeparated} we split the ZGNR$(20)$ absorption spectrum into two parts: `part I' containing only transitions involving the $1(s)$ subband, i.e., edge states, and `part II' containing the rest of the transitions. As can be seen from Fig.~\ref{fig:ZGNRvsACNTwithEdgeStateAbsorptionSeparated}, it is the latter that correlates with the tubes' absorption spectrum. Only the first absorption peak in ACNT$(21,21)$ does not have a counterpart in the ribbon spectrum. Thus, Eq.~\eqref{eq:ZGNRProblemAbsorptionPeakPositions} has a broader applicability and with its help the hidden correlation could be verified even by absorption measurements in the optical range. Equation~\eqref{eq:ZGNRProblemAbsorptionPeakPositions} describes zigzag ribbon peak positions when $j=2 \ldots w/2$ (even $w$) or $(w-1)/2$ (odd $w$).

The revealed correlation of the absorption peak positions in armchair tubes and zigzag ribbons may be affected by excitonic effects. Excitons are known to be important in one dimensional systems due to the enhanced binding energy~\cite{Loudon1959}. However, such effects rarely were a subject of investigation in the metallic families of graphene nanoribbons~\cite{Yang2008,Lu2013} and carbon nanotubes~\cite{Ando1997,Jiang2007,Hartmann2011}. Moreover, it seems that attention has never been paid to the high energy transitions, therefore this problem requires a thorough study. Yet, a general qualitative picture says that the positions of the presented peaks should be red-shifted by the amount of the binding energies. These energies can be linked to the system's transverse size by an analytical phenomenological quasi-one dimensional exciton model, which has been successfully applied to semiconducting quantum wires~\cite{Banyai1987,Ogawa1991,Ogawa1991a} and carbon nanotubes~\cite{Wang2005}. Then, since the tubes and ribbons in question have comparable widths and diameters, the binding energies and, therefore, shifts are expected to be close for both structures (neglecting the different shapes of their cross-sections), thereby preserving the unveiled correlation in the absorption spectra. Some excitonic states may require a magnetic field for their brightening if they happen to be dark ones~\cite{Shaver2007a}. We should also mention that the correlation reported here can be additionally hidden by a landscape of absorption peaks originating from $\sigma$-orbitals.

\section{\label{sec:Conclusions}Conclusions}

In summary, we considered the optical properties of zigzag graphene nanoribbons within the orthogonal $\pi$-orbital tight-binding model and effective mass approximation for polarization of the incident radiation parallel to the ribbon axis. It was analytically confirmed that the selection rules between valence and conduction subbands, $\Delta J = n-m$ is odd, and between conduction (valence) subbands only, $\Delta J = n-m$ is even,
stem from the wave function parity factor, $(-1)^J$, where $J$ is an integer numbering the energy bands. It was also shown that this parity factor originates from the ribbon's secular equation.

A comprehensive comparison of optical properties between carbon nanotubes and zigzag nanoribbons shows significant differences.
Most importantly, the concept of cutting lines~\cite{Samsonidze2003,Izumida2016}, or even its generalization to `cutting curves'~\cite{Wakabayashi2012,Deng2015b}, being unable to explain selection rules fails with respect to optical properties of zigzag graphene nanoribbons, while it works well for armchair carbon nanotubes. Nevertheless, a proper comparison reveals the absorption spectra of a zigzag nanoribbon and an armchair carbon nanotube have a correlation between the positions of the peaks originating from the $v \rightarrow c$ transitions between the bulk states , if $N_t = 2 N_r + 4$, where $N_{t,r}$ is the number of atoms in the tube's (ribbon's) unit cell, i.e. when the ribbon width is about half of the tube circumference. Putting it differently, this correlation takes place for ZGNR$(w)$ and ACNT$(w+1,w+1)$ if $w>5$. 

The analysis of the velocity operator matrix element dependencies on the electron wave vector shows that they have a smooth regular behavior at least up to $w=25$ in the whole Brillouin zone, including the Dirac ($k= \pm 2 \pi/3$) and transition ($k=k_t$) points. However, the matrix element behavior deviates significantly from the previous estimation $\sim \sin (k/2)$. For all types of transitions the magnitude of the velocity operator matrix elements attain a maximum value for $k \in \pm (\pi/2, \pi)$.

A close examination of the absorption spectra of zigzag ribbons shows they should have temperature and doping dependent absorption peaks originating from transitions between only conduction (valence) subbands, $\Delta J = 2,4, \ldots$, etc., which could be tuned, for instance, by a gate voltage. In particular, narrow zigzag ribbons with $w=6,7$ should have such prominent temperature and doping dependent absorption peaks. Although beyond the single electron tight-binding model the energy bands of zigzag ribbons are known to be modified by electron-electron interaction~\cite{Magda2014a} and the effect of the substrate, we believe that experimental observation of the tunable absorption should be possible as the latter effect, for instance, can be eliminated by system suspension. 

Finally, we point out that the obtained velocity matrix elements of single electron transitions can be utilized in further study of excitonic effects via Elliot's formula for absorption~\cite{Elliott1957,Elliott1960}.

\begin{acknowledgments}
This work was supported by the EU FP7 ITN NOTEDEV (Grant No. FP7-607521); EU H2020 RISE project CoExAN (Grant No. H2020-644076); FP7 IRSES projects CANTOR (Grant No. FP7-612285), QOCaN (Grant No. FP7-316432), InterNoM (Grant No. FP7-612624); Graphene Flagship (Grant No. 604391). The authors are very thankful to R. Keens and C. A. Downing for a careful reading of the manuscript and to A. Shytov and K. G. Batrakov for useful advice and fruitful discussions.
\end{acknowledgments}

\appendix
\section{\label{app:sec:WaveFunctionParityFactor} Wave-function parity factor}
In order to clarify the origin of the wave-function parity factor, we present in detail the simplification of the eigenvector component~\eqref{eq:ZGNRProblemEigenVectorComponent2}.

Equation~\eqref{eq:ZGNRProblemEigenVectorComponent2} can be further simplified if one expresses $2 \cos (k/2)$ in terms of the quantized momentum $\theta$ from the quantization condition~\eqref{eq:ZGNRProblemQuantizationCondition} as 
\begin{equation}
\label{eq:ZGNRProblemCosK}
2 \cos \dfrac{k}{2} = \dfrac{\sin w \theta}{\sin \left[ (w+1) \theta \right]} 
\end{equation}
and then substitutes the result into the square brackets of Eq.~\eqref{eq:ZGNRProblemEigenVectorComponent2}:
\begin{equation}
\label{eq:ZGNRProblemCosKinBrakets}
\sin p \theta - 2 \cos \dfrac{k}{2} \sin \left[ (p+1) \theta \right] = \dfrac{\sin \theta \sin \left[ (p-w) \theta \right]}{\sin \left[ (w+1) \theta \right]} \, .
\end{equation}
Note that the proper energy $E$ entering Eq.~\eqref{eq:ZGNRProblemEigenVectorComponent2} can also be re-casted only in terms of $\theta$ by substituting~\eqref{eq:ZGNRProblemCosK} into Eq.~\eqref{eq:ZGNRProblemProperEnergy}:
\begin{equation}
E (\theta) = \pm \dfrac{\gamma | \sin \theta |}{\left|\sin \left[ (w+1) \theta \right] \right|} \; .
\label{eq:ZGNRProblemProperEnergyTheta}
\end{equation}
Now making use of Eqs.~\eqref{eq:ZGNRProblemCosKinBrakets} and \eqref{eq:ZGNRProblemProperEnergyTheta}, one readily obtains that
\begin{align}
\label{eq:ZGNRProblemEigenVectorComponent21}
c_{2p+1} &=  \pm (-1)^{p+1} \dfrac{\sin \theta}{\left|\sin \theta \right|}  \dfrac{\left|\sin \left[  (w+1) \theta \right] \right|}{ \sin \left[  (w+1) \theta \right] }  \sin \left[ (p-w) \theta \right]\, ,
\end{align}
where the upper (lower) sign is applied for the conduction (valence) band state. The first ratio in the expression above is a trivial one, $\tfrac{\sin \theta}{\left|\sin \theta \right|}=1$ for $\theta \in (0,\pi)$. However, the second ratio deserves special attention because, as we will see next, it is a clue to the optical properties of zigzag ribbons.

The magnitude of the second ratio is, of course, unity, but its sign depends upon $\theta$. To determine the sign of the ratio $\tfrac{\left|\sin \left[ (w+1) \theta \right] \right|}{ \sin \left[ (w+1) \theta \right] }$ one needs to analyze it along with the quantization condition~\eqref{eq:ZGNRProblemQuantizationCondition}.  Since
absolute value is always positive the sign of the ratio is determined by the sign of its denominator defined by the secular equation solutions.

Let us investigate how secular equation solutions, $\theta_j$, are spread in the range $(0,\pi)$. For this purpose, one can continuously change the parameter $q$ from $0$ to $\infty$ similar to what is presented in Fig.~\ref{fig:ZGNRProblemTBmodelQC}. Varying $q$ between the above mentioned limits, one finds that the two values of $q$ determine the left and right ends of the intervals in each of which one $\theta_j$ is confined. By putting the parameter $q=0$ into~Eq.~\eqref{eq:ZGNRProblemQuantizationCondition}, we get $\sin w \theta_j = 0$ with $\theta_{j,\text{min}} = \pi (j-1) / w $ being solutions, while $q=\infty$ yields $\sin \left[ (w+1) \theta \right] = 0$ with $\theta_{j,\text{max}} = \pi j / (w+1)$ as solutions; in both cases $j=1 \ldots w$ enumerates solutions. It is worth noting that although the upper value of $q=2 \cos (k/2)$ is limited to $2$, we can take a greater value for an estimation because an increase of $q$ above $2$ shifts the initial interval right boundaries so that the original intervals are contained within the new $\theta$-intervals depicted in Fig.~\ref{fig:ZGNRProblemTBmodelQC}. The left boundaries of the intervals can also be pushed further left to put all the new intervals within even wider ones:
\begin{equation}
\label{eq:ZGNRProblemIntervals}
\pi (j-1)/(w+1) < \theta_j < \pi j/(w+1) \, .
\end{equation}
With inequalities~\eqref{eq:ZGNRProblemIntervals} at hand it is easy to analyze the argument of $\sin \left[ (w+1) \theta_j \right]$ for it is evident that for all $\theta_j$ satisfying inequalities~\eqref{eq:ZGNRProblemIntervals} the sine function argument $ (w+1) \theta_j $ is squeezed between $\pi (j-1)$ and $\pi j$. This leads to positive and negative signs of $\sin \left[ (w+1) \theta_j \right]$ for odd and even $j$, respectively. Therefore, the second ratio in Eq.~\eqref{eq:ZGNRProblemEigenVectorComponent21} can be written as
\begin{equation}
\dfrac{\left|\sin \left[ (w+1) \theta_j \right]\right|}{ \sin \left[ (w+1) \theta_j \right] } = (-1)^{j-1} \, ,
\label{eq:ZGNRProblemMinusOneJ}
\end{equation}
where $j$ is an integer being interpreted as the band number.

\section{\label{app:sec:EigenVectorsAtTransitionPoint} Edge and bulk state eigenvectors at the transition point}

Let us obtain the wave functions of the edge states in the explicit form and show how it reduces at the transition point $k_t$ defined as a solution of the equation $2 \cos (k/2) = w/(w+1)$. As has been mentioned above, to do this one needs to use substitution $\theta \rightarrow i \beta$, which upon application to~\eqref{eq:ZGNRProblemEigenVectorComponents4} yields
\begin{align}
\label{eq:ZGNRProblemEigenVectorComponents5}
\begin{pmatrix}
\tilde{c}^{(j)}_{2p-1} \\
\tilde{c}^{(j)}_{2p}
\end{pmatrix} &= \begin{pmatrix}
\pm i \sinh \left[ (w+1-p) \beta_j \right] \\
 i \sinh p \beta_j 
\end{pmatrix} \, ,
\end{align}
with $p=1, \ldots, w$. Note that $j=1$ for bands containing edge states, therefore the parity factor has been ruled out and $\mp$ in~\eqref{eq:ZGNRProblemEigenVectorComponents4} has been replaced with $\pm$ in~\eqref{eq:ZGNRProblemEigenVectorComponents5}. The same substitution applied to the normalization constant~\eqref{eq:ZGNRProblemTBmodelNormalizationConstant} leads to
\begin{equation}
\label{eq:ZGNRProblemTBmodelNormalizationConstant2}
N_j = \dfrac{1}{\sqrt{w-\cosh \left[  (w + 1) \beta_j \right] \, \dfrac{\sinh w \beta_j}{\sinh \beta_j}}} \, .
\end{equation}
As one can notice, the expression under the square root of~\eqref{eq:ZGNRProblemTBmodelNormalizationConstant2} is negative, therefore the imaginary unit resulting form it must cancel with that in~\eqref{eq:ZGNRProblemEigenVectorComponents5}. Hence, for normalized eigenvector components it can be written
\begin{align}
\label{eq:ZGNRProblemEigenVectorComponents6}
\begin{pmatrix}
\tilde{c}^{(j)}_{2p-1} \\
\tilde{c}^{(j)}_{2p}
\end{pmatrix} &= N_j \begin{pmatrix}
\pm \sinh \left[ (w+1-p) \beta_j \right] \\
 \sinh p \beta_j 
\end{pmatrix} \, ,
\end{align}
where $ p= 1,\dots, w$ and
\begin{equation}
\label{eq:ZGNRProblemTBmodelNormalizationConstant3}
N_j = \dfrac{1}{\sqrt{\cosh \left[ (w + 1) \beta_j \right] \, \dfrac{\sinh w \beta_j}{\sinh \beta_j}-w}} \, .
\end{equation}
Note that the eigenvector~\eqref{eq:ZGNRProblemEigenVectorComponents6} does not contain $(-1)^J$ factor like Eq.~(34) in work~\cite{Wakabayashi2012}. Even for inverse band enumeration, this factor would be $(-1)^w$ not $(-1)^J$.  At the transition point, $\beta_j \rightarrow 0$, which results in divergence in~\eqref{eq:ZGNRProblemTBmodelNormalizationConstant3} if all hyperbolic functions are expanded to the first order. However, using the original definition of the constant:
\begin{equation}
N_j = \dfrac{1}{\sqrt{ 2 \sum_{p=1}^{w} \sinh^2 p \beta_j}} \, ,
\end{equation}
where the factor of $2$ is due to the fact that $\sum_{p=1}^w \sinh p \beta_j = \sum_{p=1}^w \sinh \left[ (w+1-p) \beta_j \right]$, the same first order expansion results in
\begin{equation}
N_j = \dfrac{1}{\beta_j \sqrt{ 2 \sum_{p=1}^{w}  p^2}} \, .
\end{equation}
Thus, for normalized eigenvectors in the vicinity of the transition point, one has
\begin{align}
\label{eq:ZGNRProblemEigenVectorComponents7}
\begin{pmatrix}
\tilde{c}^{(j)}_{2p-1} \\
\tilde{c}^{(j)}_{2p}
\end{pmatrix} &= \dfrac{1}{\sqrt{2 N_c}} \begin{pmatrix}
\pm  (w+1-p) \\
  p 
\end{pmatrix} \, ,
\end{align}
where
\begin{equation}
N_c = \sum_{p=1}^w p^2 = \dfrac{w (w+1)(1+2w)}{6} \, .
\end{equation}
The same result can be obtained starting from the eigenvectors~\eqref{eq:ZGNRProblemEigenVectorComponents4} and their normalization constant specified as $N_j = 1/\sqrt{2 \sum_{p=1}^w \sin^2 p\theta_j}$, therefore wave functions approaching $k_t$ from the left and from the right attain the same value. As a result of this seamless transition of one type of functions into another, the VMEs can be obtained as smooth functions of electron wave vector $k$ for the lowest conduction (higherst valence) subbands, i.e., for $j=1$.

It is to be mentioned here that the edge states can be also obtained in zigzag carbon nanotubes with finite length~\cite{Compernolle2003}. Unlike the case of the infinite ribbon the number of such states is finite in tubes. Recently, it has been shown that this number is related to the winding number~\cite{Izumida2015,Izumida2016}. However, the state at the transition point, the charge density of which decays quadratically towards the structure center, seems to be less likely in the finite tubes.

\section{\label{app:sec:PeriodicBoundaryConditions} Periodic boundary conditions}
In this part of the appendix, we demonstrate how the fixed end (`hard wall') boundary condition employed in this paper for zigzag ribbon investigation is related to the periodic boundary condition that is used for carbon nanotubes. A carbon nanotube of the armchair type (see Ref.~\cite{SaitoBook1998} for tubes classification) is unrolled into a graphene nanoribbon with zigzag edges. The tight-binding Hamiltonian of the armchair nanotube differs from that of the zigzag ribbon by the upper right and lower left nonzero elements. For instance, for the ribbon Hamiltonian given by Eq.~\eqref{eq:ZGNRProblemHamiltonian} an equivalent tube Hamiltonian is
\begin{equation}
\label{app:ACNTHamiltonian}
    H = \begin{pmatrix}
   0 & \gamma q & 0 & \gamma  \\
\gamma q & 0 & \gamma & 0  \\
0 & \gamma & 0 & \gamma q  \\
\gamma & 0 & \gamma q & 0 
    \end{pmatrix} \, .
\end{equation}
Despite these differences the eigenproblem of such a Hamiltonian reduces to the same transfer matrix equation as Eq.~\eqref{eq:ZGNRProblemTransferMatrixEquation}. The periodic boundary condition, however, requires $C_{N+1}=C_1$, whence it follows that the secular equation is $\det(T^w-I)=0$. To obtain the explicit form of the secular equation, one can use~\eqref{eq:ZGNRProblemTw}, but there is a faster way if one uses the following relation~\cite{Kerner1954}:
\begin{equation}
\det \left( T^w - I \right) = \det T^w + \det I - \mbox{Tr} \left(T^w\right) \, .
\end{equation}
Using the above relation and taking into account that $\det T=1$, the secular equation can be recasted as
\begin{equation}
\mbox{Tr} \left(T^w\right) = 2 \, .
\end{equation}
The cyclic property of the trace operation allows further simplification of the secular equation:
\begin{equation}
\label{app:ACNTSecularEquation0}
\mbox{Tr} \left(S^{-1} \Lambda^w S \right) = \mbox{Tr} \left(\Lambda^w S S^{-1}\right) = \mbox{Tr} \left(\Lambda^w\right) = 2 \,,
\end{equation}
where $\Lambda$ is a diagonal form of the transfer matrix $T$ with the diagonal elements given by $\lambda_{1,2} = e^{\pm i \theta}$, i.e. a new variable $\theta$ is defined as $A=\cos \theta$ [cf. with Eq.~\eqref{eq:ZGNRProblemCosTheta}], $S,S^{-1}$ are given by Eqs.~\eqref{eq:ZGNRProblemSinv}. Such treatment is equivalent to that with $\lambda_{1,2}$ given by Eq.~\eqref{eq:ZGNRProblemCosTheta}, the difference is in subband enumeration similar to that mentioned for the hard wall boundary condition. In Fig.~\ref{fig:TubeVsRibbonBandStructure}, the tube's band enumeration, we refer to as direct one, corresponds to $A=-\cos \theta$. The above chosen inverse enumeration, $A=\cos \theta$, is shown in the right panel of Fig.~\ref{fig:ZGNRProblemVMEsInTBmodel}. It was chosen to obtain the tube's energy bands in a form close to graphene energy bands~\cite{Wallace1947,Saito1992,SaitoBook1998}. Thus, for an armchair tube secular equation, we end up with
\begin{equation}
\label{app:ACNTSecularEquation1}
\lambda_1^w + \lambda_2^w = 2 \cos (w \theta) = 2; \Leftrightarrow \cos (w \theta) = 1 \, ,
\end{equation}
whence it is evident that $\theta_j = 2 \pi j / w$ with $j$ being an integer numbering solutions and $w=N/2$ with $N$ being the number of carbon atoms in the tube's unit cell. To obtain the tube energy bands $\theta_j$ should be substituted into $\pm \gamma \sqrt{q^2+2q \cos \theta +1}$, which yields
\begin{equation}
\label{app:ACNTBandStructure}
E_j(k) = \pm \gamma \sqrt{4 \cos^2 \dfrac{k}{2} + 4 \cos \dfrac{k}{2} \cos \dfrac{2 \pi j}{w} + 1} \, ,
\end{equation}
where we use $j$ for the band numbering.

In the case of the hard wall boundary condition and variable $\theta$ introduced as above, i.e. with the reverse enumeration of the ribbon bands, the secular equation has the form:
\begin{equation}
\label{app:ZGNRSecularEquation}
\sin w \theta + 2 \cos \dfrac{k}{2} \sin \left[ (w+1) \theta \right] = 0 \, .
\end{equation}
The proper energy is obtained by substituting solutions of this equation into $\pm \gamma \sqrt{q^2+2q \cos \theta +1}$. Solutions of~\eqref{app:ZGNRSecularEquation} can be found in the zero approximation by setting $k = 0$; ideally, one should set $q=2 \cos (k/2) \rightarrow \infty$. This leads to $\sin \left[ (w+1) \theta \right] = 0$ with $\theta_j = \pi j/ (w + 1)$ being solution. Equating $\theta_j$ obtained for a tube and ribbon, one gets:
\begin{equation}
\label{app:NumberOfAtomsRelation}
\dfrac{2 \pi j}{N_t/2} = \dfrac{\pi j}{N_r/2 + 1} \, ,
\end{equation}
where $N_{t,r}$ is the number of atoms in the unit cell of the tube and ribbon, respectively. As follows from~\eqref{app:NumberOfAtomsRelation} if 
\begin{equation}
N_t = 2 N_r + 4 
\end{equation}
then the proper energies are approximately equal at $k=0$. It is also possible to consider the opposite limit when $k= \pi$, which leads to $\theta_j = \pi j / w$ in the case of the ribbon. The usage of this $\theta_j$ results in a better match of the ribbon and tube energies close to the edge of the Brillouin zone, i.e., at $k = \pi$, if the following relation holds between the number of atoms in the structures: $N_t = 2 N_r $.

\section{\label{app:sec:ArmchairCNTselectionRules} Armchair nanotube selection rules}
In this section, we derive selection rules for transitions in armchair carbon nanotubes (ACNTs). In spite of being known for a long time~\cite{Ajiki1994,Gruneis2003,Jiang2004,Goupalov2005,Malic2006,Zarifi2009}, they have not been derived from the full tight-binding Hamiltonian. The purpose of this exercise is to provide deeper understanding of the difference in the optical properties of zigzag graphene nanoribbons and ACNTs and also to show their relation to the graphene single layer sheet.

To calculate velocity operator matrix elements, one needs the wave functions. 
Substitution of Eq.~\eqref{app:ACNTSecularEquation1} solution $\theta_j = \frac{2 \pi j}{w}$ into $T^w-I$ gives a zero matrix. Hence the boundary condition $C_{N+1} = C_1; \rightarrow (T^{w}-I) C_1 = 0$ is fulfilled for any components of the initial vector $C_1$. We see that for the periodic boundary condition the initial vector $C_1$ can be an arbitrary one.
The most reasonable choice of $C_1$ is one of the eigenvectors~\eqref{eq:ZGNRProblemV1andV2}. Let it be $V_2$. Then, with $\lambda_{1,2}=e^{\pm i \theta}$ the wave-function components can be found from Eq.~\eqref{eq:ZGNRProblemTransferMatrixEquation} as follows:
\begin{align}
    c^{(j)}_{2p-1} &= \pm e^{-i \theta_j (p - 1)} \dfrac{f_j}{|f_j|}, & c^{(j)}_{2p} &= e^{-i \theta_j p} \,, \label{app:sec:ACNTproblemEigenVectorComponents0}
\end{align}
where $p=1, \ldots, w$, $f_j=1+q e^{- i \theta_j}$, and we have changed the order of the components as it was done for Eq.~\eqref{eq:ZGNRProblemEigenVectorComponent22}. Introducing new function $\tilde{f}_j = e^{i \theta_j /3} f_j$ into Eq.~\eqref{app:sec:ACNTproblemEigenVectorComponents0} and applying the unitary transform $U_j = \{u_{2p -1,2p -1}, u_{2p, 2p}\}= \{e^{i \theta_j (p - 2/3)},e^{i \theta_j p}\}|_{p=1\ldots w} $ to the vector $\left| c^{(j)} \right\rangle$, we obtain
\begin{align}
    \tilde{c}^{(j)}_{2p-1} &= \pm \dfrac{\tilde{f}_j}{|\tilde{f}_j|}; &
    \tilde{c}^{(j)}_{2p} &= 1 \,,
    \label{app:sec:ACNTproblemEigenVectorComponents1}
\end{align}
where $p=1, \ldots, w$. The normalization constant $N_j = 1 / \sqrt{2 w}$ for $\left| \tilde{c}^{(j)} \right\rangle$ and it is independent of $\theta_j$.

As one can see, the unitary matrix $U_j$ depends on the band index $j$, therefore the new Hamiltonian that preserves the matrix element upon the transfromation of the $\left| c^{(n,m)} \right\rangle$ vectors is $\tilde{H} = U_{n} H U_{m}^{\dagger}$. However, such a Hamiltonian satisfies the time independent Schrodinger equation only if $n=m$. This is the selection rule for ACNT optical transitions, which also means all transitions $c \rightarrow c$ and $v \rightarrow v$ are forbidden.

For $\tilde{H} = U_j H U^{\dagger}_j$ the components of the vectors $\left| \tilde{\zeta}^{(j)} \right\rangle$ are
\begin{align}
\label{app:ACNTProblemZetaVectorComponents}
\tilde{\zeta}^{(j)}_{2p-1} &=  - \dfrac{\gamma a}{\hbar} \sin (\dfrac{k}{2}) e^{-2 i \theta_j / 3} \,, \qquad p=1,\dots, w \, ;   \nonumber \\
\tilde{\zeta}^{(j)}_{2p} &= \mp \dfrac{\gamma a}{\hbar} \sin (\dfrac{k}{2})  e^{-2 i \theta_j / 3} \dfrac{\tilde{f}_j}{|\tilde{f}_j|} \, ,  
\end{align}
with the upper ``$-$'' ( lower ``$+$'') being used for the conduction (valence) subbands. By putting Eqs.~\eqref{app:sec:ACNTproblemEigenVectorComponents1} and~ \eqref{app:ACNTProblemZetaVectorComponents} into Eq.~\eqref{eq:OpticalMatrixElementsVME}, and accounting for the normalization constant $N_j$, for allowed transitions we have
\begin{align}
\label{app:ACNTVMETBmodel}
M_{n(c),n(v)} &= -\dfrac{\gamma a}{\hbar} \sin (\dfrac{k}{2}) \dfrac{\tilde{f}_n^{\ast} e^{-2i\theta_j/3} - \tilde{f}_n e^{2 i \theta_n/3}}{2 |\tilde{f}_n|}  \, ,  \nonumber \\
&= \dfrac{\gamma a}{\hbar} \dfrac{\tilde{f}_n^{\ast} (d \tilde{f}_n/d k) - \tilde{f}_n (d \tilde{f}^{\ast}_n/d k)}{2 |\tilde{f}_n|} \, .
\end{align}
Similarly, calculations for the group velocity yields
\begin{equation}
\label{app:ACNTgroupvelocityTBmodel}
M_{n(s),n(s)} = \pm \dfrac{\gamma a}{\hbar} \dfrac{\tilde{f}_n^{\ast} (d \tilde{f}_n/d k) + \tilde{f}_n (d \tilde{f}^{\ast}_n/d k)}{2 |\tilde{f}_n|}
\end{equation}
where ``$+$" (``$-$") refers to the conduction (valence) subbands.

The same result is obtained from the graphene Hamiltonian and eigenvectors: $ (1/\hbar) \left\langle c_c \right| \partial H / \partial k_y \left| c_v \right\rangle$  with $H_{11}=H_{22}=0$, $H_{12} = H^{\ast}_{21} = \gamma \left( e^{i k_x a / \sqrt{3}} + 2 e^{-i k_x a / 2 \sqrt{3}} \cos (k_y a / 2) \right)$ and $k_x = 2 \pi j / C_h$, where $C_h$ is the tube circumference and $a=2.46$~{\AA} is the graphene lattice constant.
If $\theta_j = \sqrt{3} k_x a / 2$, $k = k_y a$, and the tube chiral index is $w/2$, then $k_x = 4 \pi j /(\sqrt{3} a w) = 2 \pi j / C_h$. Hence, Eq.~\eqref{app:ACNTVMETBmodel} can be restored by cutting graphene's optical transition matrix elements along the lines specified by the quantization of $k_x$. Finally, we note that a calculation of the matrix elements with the eigenvectors~\eqref{app:sec:ACNTproblemEigenVectorComponents0} and the Hamiltonian~\eqref{app:ACNTHamiltonian} also provides straightforward justification of the selection rules for it results in zero matrix elements when $n \neq m$.

\section{\label{app:sec:SuplementaryResults} Supplementary results}
\begin{figure}[t]
      \includegraphics[width=0.47\textwidth]{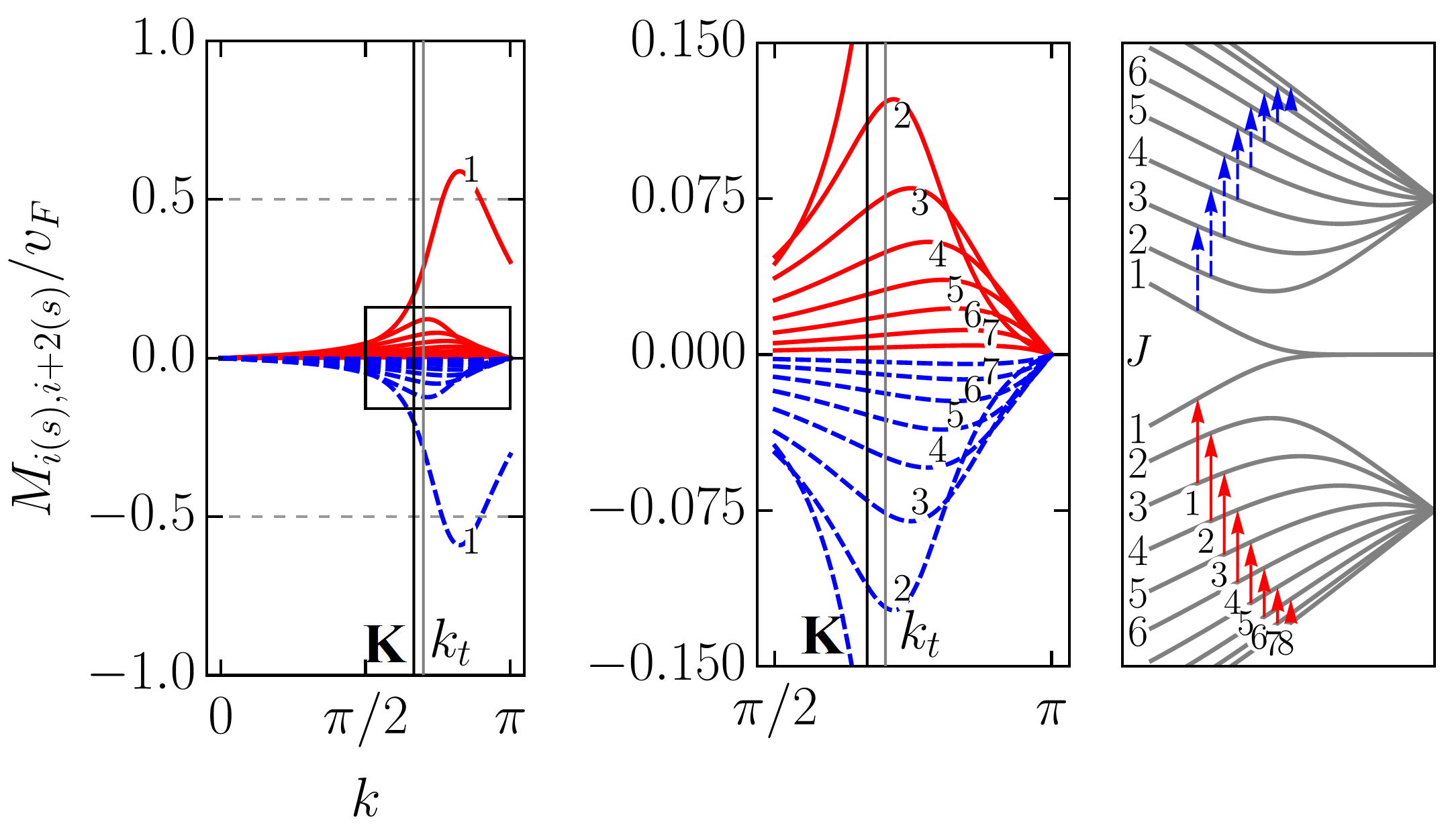} 
	\caption{The same as~Fig.~\ref{fig:ZGNRProblemVMEsInTBmodel2} (b) but for transitions between valence (conduction) subbands of lower (higher) energy: $v \rightarrow v; c \rightarrow c; \Delta J = 2$. As the plot is symmetric with respect to $k=0$, only half of the BZ is presented. The part of the plot denoted by a rectangle is zoomed in the right panel followed by the transition scheme. The VME curves correspond to the transitions labeled with the same number in the scheme.}
	\label{fig:ZGNRProblemVMEsInTBModel3}
\end{figure}

For the sake of completeness, in Fig.~\ref{fig:ZGNRProblemVMEsInTBModel3}, we present VME curves obtained for transitions between the lower (higher) energy valence (conduction) subbands.
These transitions can be referred to as $j(s) \rightarrow (j+2)(s)$, where $j=1,\ldots,w-2$. Noticing that the curve labeled by $1$ in Fig.~\ref{fig:ZGNRProblemVMEsInTBModel3} is the same as the curve labeled by $2$ in Fig.~\ref{fig:ZGNRProblemVMEsInTBmodel2} (b), one easily sees that the transitions labeled from $2$ to $7$ are much weaker compared to the transitions in Fig.~\ref{fig:ZGNRProblemVMEsInTBmodel2}. Unlike the VME curves in Figs.~\ref{fig:ZGNRProblemVMEsInTBmodel} (a) and~\ref{fig:ZGNRProblemVMEsInTBmodel2}, all curves of $j(s) \rightarrow (j+2)(s)$ transitions converge to zero at the edge of the BZ and have extrema decreasing in magnitude and shifting from the {\bf K}({\bf K}$^{\prime}$) point towards the BZ edge for greater $j$'s.

\begin{figure}[t]
      \includegraphics[width=0.47\textwidth]{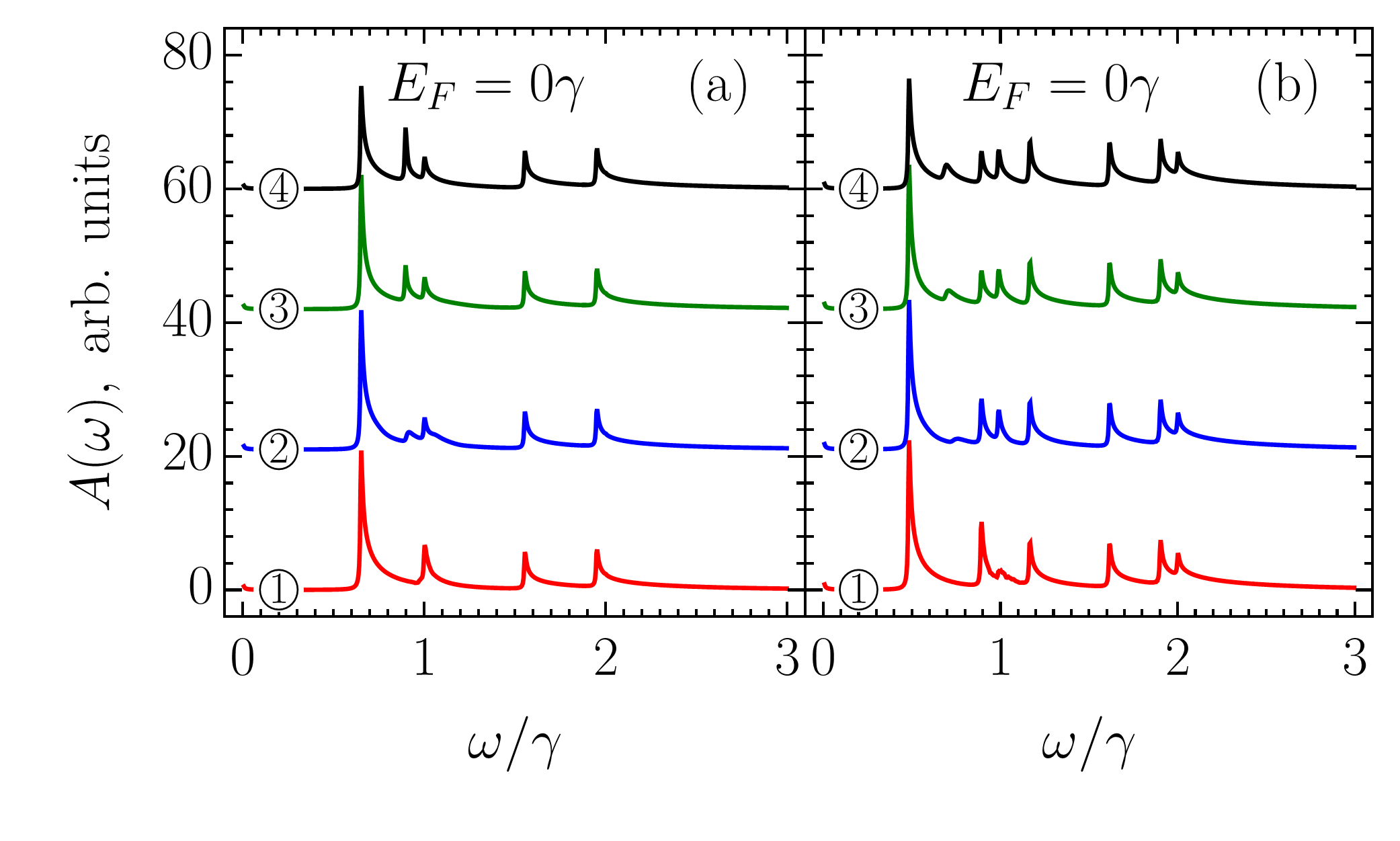}
	\caption{The absorption spectra of zigzag ribbons with (a) $w=6$ and (b) $w=9$ for different temperatures: $T = 0$, $4$, $77$, and $300$~K$/\gamma$ for curves \protect\textcircled{\raisebox{-1.2pt}{1}}, \protect\textcircled{\raisebox{-1.2pt}{2}}, \protect\textcircled{\raisebox{-1.2pt}{3}}, and \protect\textcircled{\raisebox{-1.2pt}{4}}, respectively. Absorption spectra are shifted vertically for clarity.}
	\label{fig:ZGNRProblemSpectralFunctionTdependence}
\end{figure}

Figure~\ref{fig:ZGNRProblemSpectralFunctionTdependence} shows that temperature has a similar influence on the absorption spectra to doping. The observed changes are explained in the same way as presented for Fig.~\ref{fig:ZGNRProblemSpectralFunctionFLdependence}. The peak due to the transitions $1(c)\rightarrow 3(c)$ is weaker and broader for ZGNR($9$) compared to that in ZGNR($6$). At the same time, the peak at $\omega = \gamma$ due to transitions $1(c) \rightarrow 5(c)$ is quite intense.

\bibliography{library}

\end{document}